\documentclass[preprint,showpacs,aps,preprintnumbers,amsmath,amssymb]{revtex4-1}

\usepackage{graphicx,epsfig}% Include figure files
\usepackage{dcolumn}% Align table columns on decimal point
\usepackage{bm}% bold math
\usepackage{xcolor}
\newcommand{\be}{\begin{equation}}
\newcommand{\ee}{\end{equation}}
\newcommand{\bse}{\begin{subequations}}
\newcommand{\ese}{\end{subequations}}
\newcommand{\bary}{\begin{eqnarray}}
\newcommand{\eary}{\end{eqnarray}}
\newcommand{\bwt}{\begin{widetext}}
\newcommand{\ewt}{\end{widetext}}

\begin{document}

%\preprint{ICN/000-HEP}

\title{Multi-TeV flaring  in nearby High Energy Blazars: A photohadronic scenario}
% Force line breaks with \\
\author{Sarira Sahu}
\email{sarira@nucleares.unam.mx}
\affiliation{Instituto de Ciencias Nucleares, Universidad Nacional Aut\'onoma de M\'exico, 
Circuito Exterior, C.U., A. Postal 70-543, 04510 Mexico DF, Mexico}

%\affiliation{Instituto de Ciencias Nucleares, Universidad Nacional Aut\'onoma de M\'exico, 
%Circuito Exterior, C.U., A. Postal 70-543, 04510 Mexico DF, Mexico}

%\affiliation{$^{b}$Astrophysical Big Bang Laboratory, RIKEN, Hirosawa, Wako, Saitama 351-0198, Japan}

%\ead{sarira@nucleares.unam.mx}

\begin{abstract}

Blazars are a subclass of AGN and flaring in multi-TeV gamma-ray seems to be the
major activity in high energy blazars a subgrup of blazars. Flaring is also unpredictable and switches between quiescent and
active states involving different time scales and fluxes. 
While in some high energy blazars a strong temporal correlation
between X-ray and multi-TeV gamma-ray has been observed,
outbursts in some other have no low energy counterparts and explanation of such extreme activity
needs to be addressed through different mechanisms as it is not
understood well. The extragalactic background light (EBL) plays an important role in
the observation of these high energy gamma-rays as it attenuates through
pair production of electron-positron and also changes the spectral
shape of the high energy photons.
In the context of the photohadronic model and taking EBL correction into
account, flaring can be explained very well. In a series of papers we
have developed this model to explain multi-TeV flaring events
form many  blazars. Here in this review, the 
photohadronic model is discussed and
applied to explain the multi-TeV flaring from nearby high energy 
blazars: Markarian 421, Markarian 501 and 1ES1959+650. 

\end{abstract}

%\pacs{98.54.Cm; 98.70.Rz; 98.70.Sa}% PACS, the Physics and Astronomy                     
                       % Classification Scheme.
\keywords{Blazar, TeV flare, gamma-ray}

\maketitle

\section{Introduction}
\label{sec:intro}

Blazars are a
subclass of AGN and the dominant extra galactic population in
gamma-rays\cite{Acciari:2010aa}. These objects show rapid variability
in the entire 
electromagnetic spectrum and have non thermal spectra which implies
that the observed photons originate within the highly relativistic
jets oriented very close to the observers line of
sight\cite{Urry:1995mg}.
Due to the small viewing angle of the jet, it is possible to observe the
strong relativistic effects, such as the boosting of the emitted power
and a shortening of the characteristic time scales, as short as
minutes\cite{Abdo:2009wu,Aharonian:2007ig}. Thus these objects are important to study the energy
extraction mechanisms from the central super-massive black hole, 
physical properties of the astrophysical jets,
acceleration mechanisms of the charged particles in the jet and production of
ultra high energy cosmic rays, very high energy $\gamma$-rays and neutrinos.

The spectral energy distribution (SED) of these blazars
has a double peak structure in the $\nu-\nu F_{\nu}$ plane. 
The low  energy peak corresponds to
the synchrotron radiation from a population of relativistic electrons
in the jet and the high energy
peak believed to be due to the synchrotron self
Compton (SSC) scattering of the high energy electrons with their
self-produced synchrotron photons\cite{Dermer:1993cz,Sikora:1994zb}. 
Depending mostly on the optical spectra, blazars can be divied into 
BL Lacertae objects (BL Lacs) and flat- spectrum radio quasars
(FSRQ)\cite{Padovani:2015mba}. Based on the  location of the first peak, BL Lacs can
be further classified into  low energy peaked
blazars (LBLs) ($\nu_{syn}^{peak} < 10^{14}\, Hz$), intermediate energy
peaked blazars  (IBLs) ($10^{14}\, Hz < \nu_{syn}^{peak} < 10^{15}\, Hz$) and high energy
peaked blazars (HBLs) ($\nu_{syn}^{peak} > 10^{15}\, Hz$)\cite{Padovani:1994sh}. 
As leptons ($e^{\pm}$) are responsible for the
production of the SED, this is called the {\it leptonic model} and in general is
very successful in explaining the multiwavelength emission from
blazars and FR~I
galaxies\cite{Fossati:1998zn,Ghisellini:1998it,Abdo:2010fk,Roustazadeh:2011zz}.
The inevitable outcome of the leptonic models is that,
flaring at TeV energy should be accompanied by a simultaneous flaring
in the synchrotron peak. However, non observation/suppression of low energy counterparts have been observed in
many flaring blazars, for example, flare of May 2002 from the HBL 
1ES 1959+650\cite{Holder:2002ru,Aharonian:2003be,Krawczynski:2003fq}
and 2004 flare from Markarian 421\cite{Blazejowski:2005ih}. 
These observations are
in favor of hadronic model and/or hybrid (hadronic + leptonic) models. 
It is to be noted that, recent observation of  the high energy
neutrino event by IceCube neutrino observatory is correlated with a
flaring blazar \cite{IceCube:2018cha,Ahnen:2018mvi}. So very high
energy (VHE) protons/nuclei should be produced by
the blazar so that their interaction with the surrounding background
can produce pions and subsequent decay of pions will produce neutrinos
and gamma-rays. This shows that hadronic models play important role
here.

Flaring seems to be the major activity of the blazars
which is unpredictable and switches between quiescent and
active states involving different time scales and fluxes. 
While in some blazars a strong temporal correlation
between X-ray and multi-TeV $\gamma$-ray has been observed,
outbursts in some other have no low energy counterparts (orphan
flaring)\cite{Krawczynski:2003fq,Blazejowski:2005ih} and explanation of such extreme activity
needs to be addressed through different mechanisms. It is also very
important to have simultaneous multiwavelength observations of the
flaring period to constrain different theoretical models of emission
in different energy regimes. 

%\section{Models to explain blazar flaring}

Different theoretical models have been proposed to explain the flaring
from AGN and its subclasses. These models are mainly clssified into
two categories: the leptonic models and the hadronic models. In the
leptonic model scenario, the high energy electrons upscatter the low
energy photons in the jet through the SSC process and the photons can
be in the GeV-TeV region. This model has severe limitations to explain
VHE $\gamma$-rays and the orphan flaring observed in mnay blazars. The
multi-zone leptonic model can explain these
high energy emissions, however, one has to increase the number of
parameters in the model.
In the hadronic synchrotron-proton blazar model\cite{Muecke:2002bi,Reimer:2004a,Aharonian:2000pv},  emission of synchrotron
photons from protons take place. In this scenario, 
the Fermi accelerated protons in the jet magnetic field, emit synchrotron
radiation which will be suppressed by a factor of
$m_p^{-4}$, where $m_p$ is the proton mass. So ultra high energy
proton flux is needed to explain the VHE gamma-rays. It also needs a
strong magnetic field for the synchrotron process to be effective, but
a strong magnetic field in the jet is not very usual.
In the jet-in-jet model of Giannios et al. \cite{Giannios:2009pi} minijets are formed
within the jet due to flow instabilities and these minijets move
relativistically with respect to the main jet flow. The interaction of
the daughter jets with the main jet are responsible for the production
of VHE gamma rays. While the minijets are aligned with our line of
sight, the VHE gamma rays are beamed with large Doppler factor.
This scenario can explain the 2010 flare of the radio galaxy M87 but does not provide a
quantitive prediction of the light curve of the flare.
The lepto-hadronic model\cite{Reynoso:2010pp} fits to the low energy
$\gamma$-ray spectrum by Fermi/LAT and HESS low state but not the flaring state.
Similarly, the magnetosphere model \cite{Rieger:2007tt,Levinson:2010fc,Rieger:2011ch} 
can explain the hard TeV
spectrum but in this case also there is no detailed quantitive predication for the
VHE light curve.
Also interaction of Fermi acclerated protons with the MeV photons
emitted by the Wein fireball in the base of the jet can explain the
orphan falring from 1ES 1959+650 and Mrk
421\cite{Fraija:2015xha}. It is well
known that EBL plays an inportant role for
the attenuation of $\gamma$-rays even from the nearby blazars and this
model does not take into account the EBL effect. The VHE SED of 
blazars are modelled using different hadronic models\cite{Petropoulou:2016sll,Petropoulou:2016xat,Rachen:1999dq,Aharonian:2001cp}. Using the
hadronic models, the high energy cosmic ray and the neutrino fluxes
are also estimated\cite{Halzen:2005pz,Fraija:2016yeh,Petropoulou:2016ujj}.

In a series of papers  Sahu et
al. \cite{Sahu:2013ixa,Sahu:2012wv,Sahu:2013cja,Sahu:2015tua,Sahu:2016mww,Sahu:2016bdu,Sahu:2018rqs,Sahu:2018gik}
have explained
the GeV-TeV flaring from many blazars using photohadronic
scenario. In this scenario, Fermi accelerated protons interact with
the background photons in the jet environment to produce
$\Delta$-resonance. Subsequent decay of the $\Delta$-resonance to
pions can produce VHE photons and neutrinos. The produced gamma-rays
can explain very well the observed spectra from the flaring blazars.

The TeV photons of the flare can interact with the background soft
photons in the jet to produce $e^+e^-$
pairs. However, production of the lepton pair within the jet depends on the 
size of the emitting region and the photon density in it. Also the  
required target soft photon threshold 
energy $\epsilon_{\gamma} \ge 2 m^2_e/E_{\gamma}$ is needed. 
It has been observed that the jet medium is transparent to pair production where
the optical depth is very small\cite{Sahu:2013ixa,Abdop:2010}.
Also the TeV photons on their way to Earth can interact with  the extragalactic
background light (EBL) to produce the lepton 
pair\cite{Stecker:1992wi,Hauser:2001xs,Dominguez:2010bv,Franceschini:2008tp}. TeV
photons from the sources in the cosmologically local Universe (low redshift
sources) are believed to propagate unimpeded by the EBL, 
although the effect is found to be non negligible\cite{Abdop:2010}.

In this work the goal is to use the photohadronic model of Sahu et al.\cite{Sahu:2013ixa} 
and different template EBL
models\cite{Franceschini:2008tp,Inoue:2012bk} to explain the observed
GeV-TeV flares from different HBLs. 
In this article I review the photohadronic model of Sahu et al.  and discuss about its
applicability to explain the flaring events from high energy
blazars. The plan of the papers is as follows. In Sec. II a short 
account of different EBL models is given. In Sec. III 
a detail discussion about the photohadronic model and the kinematical condition for
the process $p\gamma\rightarrow \Delta$ and its subsequent decay is
given. Also the relation between the observed GeV-TeV
gamma-ray flux and the gamma-ray from the photohadronic process is shown.
Sec. IV is dedicated to discuss the results of different
flaring blazars and compare the results with other models. 
In Sec. V a general remark is given regarding the problems of
different models. Finally I
summarize the photohadronic model and discuss about the future
research in this field in Sec. VI.

\section{EBL Models}
\label{sec:eblmodels}
A distinctive feature of the high energy $\gamma$-ray astronomy is that 
%The extragalactic very high energy (VHE, $E_{\gamma} > 100$ GeV)
the high energy gamma-rays undergo
energy dependent attenuation en route to
Earth by the intervening EBL
through electron-positron pair production\cite{Stecker:1992wi}. This interaction process
not only attenuates the absolute flux but also significantly 
changes the spectral shape of the VHE photons.
%changes the VHE emission spectrum. 
The diffuse EBL contains a record of the star
formation history of the Universe. A proper 
understanding of the EBL SED
is very important for the correct interpretation of the
deabsorbed VHE spectrum from the source.
The direct measurement of the EBL is very difficult with
high uncertainties mainly due to the contribution of
zodiacal light\cite{Hauser:2001xs,Chary:2010dc}, and
galaxy counts result in a lower limit since the number of unresolved
sources (faint galaxies) is unknown\cite{Madau-Po:2000}.

Keeping in mind
the observational constraints and the uncertainty associated with the direct detection of
the EBL contribution, several approaches with different
degrees of complexity have been developed to
calculate the EBL density as a function of energy for different
redshifts. A wide range of models have been developed to model
the EBL SED based on our knowledge of galaxy and star formation rate
and at the same time incorporating the observational
inputs\cite{Dominguez:2010bv,Franceschini:2008tp,Inoue:2012bk,Kneiske:2002wi,Stecker:2005qs,Orr:2011ha,Primack:2005rf}. Mainly
three types of EBL models exist: backward and forward evolution models
and semi-analytical galaxy formation models with a combination of
information about galaxy evolution and
observed properties of galaxy spectra.
In the backward evolution scenarios\cite{Stecker:2005qs},  one starts from the observed
properties of galaxies in the local universe and evolve them from cosmological initial conditions or
extrapolating backward in time using parametric models of the
evolution of galaxies. This extrapolation induces
uncertainties  in the properties of the EBL which increases at high redshifts.
However, the forward evolution
models\cite{Franceschini:2008tp,Kneiske:2002wi} predict the temporal
evolution of galaxies forward in time starting from the cosmological
initial conditions. Although, these models are successful in
reproducing the general characteristics of the observed EBL, cannot
account for the detailed evolution of  important quantities such as
the metallicity and dust content, which can significantly affect the
shape of the EBL. Finally, semi-analytical models have been developed
which follow the formation of large scale structures driven by cold dark
matter in the universe by using the cosmological parameters from
observations. This method also accounts for the merging of the dark matter
halos and the emergence of galaxies which form as baryonic matter falls into the
potential wells of these halos. Such models are successful in
reproducing observed properties of galaxies from local universe up to
$z\sim 6$.

%%%%%%%%%%%%%%%%%%%%%%%
\begin{figure}[t!]%fig1
%\vspace{0.3cm}
{\centering
\resizebox*{0.8\textwidth}{0.6\textheight}
%{\includegraphics{Flaring-Model1.pdf}}
{\includegraphics{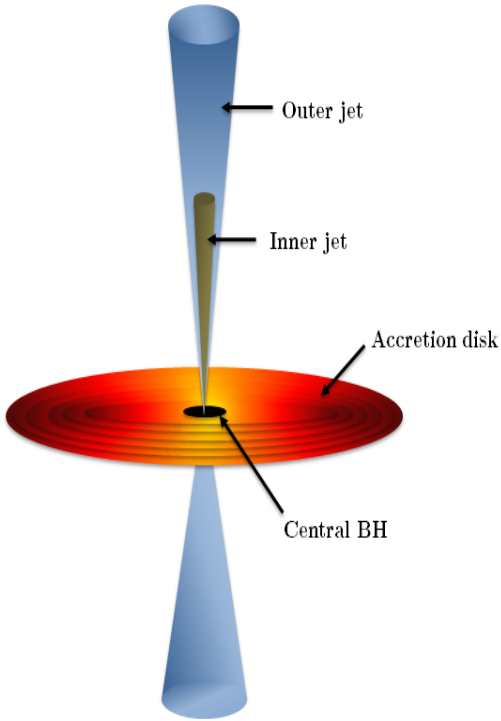}}
\par}
\caption{Geometry of the flaring of HBL: the interior
  compact cone (jet) is responsible for the flaring event and the exterior cone
  corresponds to the normal jet. 
%(See the electronic edition of the Journal for a color version of this figure).
}
\label{blaz:fig1}
\end{figure}
%%%%%%%%%%%%%%%%%%%%%%%%%%%

\section{Photohadronic Scenario}
In general, in the leptonic one-zone synchrotron and SSC jet  model
the emitting region is a blob
with comoving radius $R'_b$ (where $^{\prime}$ implies the jet
comoving frame and without prime are in the observer frame),  moving with a velocity
$\beta_c$ corresponding to a bulk Lorentz
factor $\Gamma$ and seen at
an angle $\theta_{ob}$ by an observer which results with a Doppler
factor ${\cal D}=\Gamma^{-1} (1-\beta_c
\cos\theta_{ob})^{-1}$\cite{Ghisellini:1998it,Krawczynski:2003fq}.  
The emitting region is filled with an
isotropic electron population and a randomly oriented magnetic field
$B'$. The electrons have a power-law spectrum.
The energy spectrum of the Fermi-accelerated protons in the blazar jet
is also assumed to be of power-law. Due to high radiative losses, electron
acceleration is limited. On the other hand, protons and heavy nuclei can
reach UHE through the same acceleration mechanism. 

The photohadronic model discussed here relies on
the above standard interpretation of the leptonic model to explain both low and
high energy peaks by synchrotron and SSC photons respectively as in the
case of any other AGN. Thereafter, it is assumed that the
flaring occurs within a compact and confined volume of size $R'_f$
inside the blob of radius $R'_b$. The geometrical description of the jet structure
during a flare is shown in Fig. \ref{blaz:fig1}.
In this scenario the internal and the external jets are moving with
almost the
same bulk Lorentz factor $\Gamma_{in}\simeq \Gamma_{ext}\simeq \Gamma$
and the Doppler factor ${\cal  D}$ as the blob (for blazars
$\Gamma\simeq {\cal  D}$). 
Within the confined volume, the injected spectrum of the
Fermi accelerated charged particles have 
a power-law, and for the
protons with energy $E_p$ it is given as
\be
\frac{dN_p}{dE_p}\propto  E_p^{-\alpha},
\label{powerlaw}
\ee
where the spectral index $\alpha \ge 2$\cite{Dermer:1993cz}. Also in this small volume, the comoving
photon number density 
$n'_{\gamma, f}$ (flaring) is much higher than rest of the blob
$n'_{\gamma}$ (non-flaring).

The dominant mechanism through which the high energy protons interact
with the background photons in the inner jet region is given by
\be
p+\gamma \rightarrow \Delta^+\rightarrow  
 \left\{ 
\begin{array}{l l}
 p\,\pi^0, & \quad \text {fraction~ 2/3}\\
  n\,\pi^+ , 
& \quad  \text {fraction~ 1/3}\\
\end{array} \right. ,
\label{decaymode}
\ee
which has a cross section $\sigma_{\Delta}\sim 5\times 10^{-28}\,
{\rm cm}^2$. Subsequently, the charged and neutral pions  will
decay through
$\pi^+\rightarrow e^+{\nu}_e\nu_{\mu}{\bar\nu}_{\mu}$ and
$\pi^0\rightarrow\gamma\gamma$ respectively. 
The inner compact region has a photon density much higher than the
outer region. Due to the adiabatic expansion of the inner jet, the 
photon density will decrease when it crosses into the outer region. 
We assume the scaling behavior of the photon densities in
the inner and the outer regions. Mathematically we can express this as
\be
\frac{n'_{\gamma, f}(\epsilon_{\gamma_1})}
{n'_{\gamma, f}(\epsilon_{\gamma_2})} \simeq \frac{n'_\gamma(\epsilon_{\gamma_1})}
{n'_\gamma(\epsilon_{\gamma_2})},
\label{densityratio}
\ee 
i.e. the ratio of photon densities at two different
background energies $\epsilon_{\gamma_1} $  and $\epsilon_{\gamma_2} $
in the flaring ($n'_{\gamma, f}$) and in the non-flaring ($n'_{\gamma}$)
states remains almost the same. The photon density in the outer region
is known to us from the observed flux. So by using the above relation we can express
the unknown inner photon density in terms of the known outer
density which can be calculated from the observed flux in the usual
way using the observed/fitted spectral energy distribution
(SED). Henceforth, for our calculation, we shall use $n'_{\gamma}$ and its corresponding
flux rather than the one from the inner jet region which is not known.

\subsection{Kinematical condition}

For the above process in Eq.(\ref{decaymode}) to take place, the center-of-mass energy of the
interaction has to exceed the $\Delta$-mass 1.232 GeV which  corresponds
to the kinematical condition
\be
E'_p \epsilon'_\gamma= \frac{(m^2_{\Delta}-m^2_p)} {2 (1-\beta_p
  \cos\theta)}
\simeq {0.32\,{\rm  GeV}^2},
\ee
where $E'_p$ and $\epsilon'_\gamma$ are
the proton and the background  photon energies in the comoving frame
of the jet, respectively. Also for high energy protons we take
$\beta_p\simeq 1$. Since in the comoving frame the protons collide
with the SSC photons
from all directions, in our calculation we consider an average value $(1-\cos\theta) \sim 1$ 
($\theta$ in the range of 0 and $\pi$). 
In the observer frame, one can
re-write the matching condition as
\be
E_p \epsilon_\gamma \simeq 0.32~ \frac{\Gamma {\cal D}}{(1+z)^2} ~{\rm GeV}^2~.
\label{resonant1} 
\ee
Here 
\be
\epsilon_\gamma = \frac{{\cal D} \epsilon'_\gamma}{(1+z)},
\ee
is the observed background photon energy, while
\be
E_p=\frac{\Gamma E'_p}{(1+z)}, 
\ee
is the energy of the proton as measured by the observer on Earth,
if it could escape the source and reach earth without energy loss and
$z$ is the redshift of the object. 

In the comoving frame, each pion carries $\sim 0.2$ of the proton energy.
Considering that each $\pi^0$ decays into two $\gamma$-rays, the $\pi^0$-decay
$\gamma$-ray energy in the observer frame ($E_{\gamma}$)can be written as
\be
E_\gamma = \frac{1}{10}\frac{{\cal D}}{(1+z)} E'_p = \frac{\cal
  D}{10\,\Gamma} E_p. 
\label{EgammaEp}
\ee
The matching condition between the $\pi^0$-decay photon energy
$E_{\gamma}$ and the 
target photon energy $\epsilon_{\gamma}$ is therefore
\be
E_\gamma \epsilon_\gamma \simeq 0.032~\frac{ {\cal D} \Gamma}{(1+z)^2} ~{\rm GeV}^2.
\label{Eegamma} 
\ee
So from the known flare energy $E_{\gamma}$ of a blazar the seed
photon energy $\epsilon_{\gamma}$ can be calculated when the Doppler
factor and the bulk Lorentz factor is known from the leptonic model
fit to the blazar SED.

\subsection{Flux calculation}

The observed VHE $\gamma$-ray flux depends on the background seed
photon density and the differential power-spectrum of the Fermi
accelerated protons given as
$F_{\gamma}\propto n'_{\gamma} (E^2_p\,dN/dE_p)$.
It is to be noted that, the photohadronic process in a standard
blazar jet environment is inefficient due to low seed photon
density $n'_{\gamma}$. So to explain the multi-TeV emission from the
flaring in the photohadronic scenario, 
jet kinetic power has to be increased to the super-Eddington limit\cite{Cao:2014nia,Zdziarski:2015rsa}.
However, the inner compact jet scenario evades this problem due to the
higher photon density\cite{Sahu:2013ixa}.

The optical depth of the $\Delta$-resonance process in the inner jet region
is given by
\be
\tau_{p\gamma}=n'_{\gamma, f} \sigma_{\Delta} R'_f.
\label{optdepth}
\ee
The efficiency of the  $p\gamma$ process depends on the
physical conditions of the interaction region, such as the size, 
the distance from the base of the jet, the photon density and their
distribution in the region of interest.

In the inner region we compare the dynamical time scale $t'_{d}=R'_f$ 
with the $p\gamma$ interaction time scale
$t'_{p\gamma}=(n'_{\gamma,f}\sigma_{\Delta} K_{p\gamma})^{-1}$ to
constrain the seed photon density so that multi-TeV photons can be
produced. For a moderate efficiency of this process, we can assume
$t'_{p\gamma} > t'_{d}$ and this gives
$\tau_{p\gamma} < 2$, where the inelasticity parameter is assigned with the
usual value of 
$K_{p\gamma}=0.5$. 
Also by assuming the Eddington luminosity is equally shared by the jet
and the counter jet, the luminosity within the inner region for a seed
photon energy $\epsilon'_{\gamma}$ will satisfy $(4\pi n'_{\gamma,f}
R'_f \epsilon'_{\gamma}) \ll L_{Edd}/2$.  This puts an upper limit on
the seed photon density as 
\be
n'_{\gamma,f}\ll \frac{L_{Edd}} {8\pi R'^2_f
\epsilon'_{\gamma}}.
\label{nedd}
\ee
From Eq.(\ref{nedd})  we can estimate the photon density in this
region. In terms of SSC photon energy and its luminosity, the photon number density
$n^{\prime}_{\gamma}$ is expressed as
\be
n^{\prime}_{\gamma}(\epsilon_{\gamma}) = \eta \frac{L_{\gamma,
    SSC} (1+z)}{{\cal D}^{2+\kappa} 4\pi {R^{\prime}}^2_b
  \epsilon_{\gamma}},
\label{nblob}
\ee
where $\eta$ is the efficiency of SSC process and $\kappa$ describes
whether the jet is continuous ($\kappa=0$) or discrete ($\kappa=1$). In this
work we take $\eta=1$ for 100\% efficiency.
The SSC photon luminosity  is expressed in terms of the observed flux
 ($\Phi_{SSC}(\epsilon_{\gamma}) =\epsilon^2_{\gamma} dN_{\gamma}/d\epsilon_{\gamma}$) and is
given by
\be
L_{\gamma,SSC}=\frac{4\pi d^2_L
  \Phi_{SSC}(\epsilon_{\gamma})}{(1+z)^2}.
\label{lssc}
\ee
Using the Eqs. (\ref{nblob}) and (\ref{lssc})  we can simplify the
ratio of photon densities given in Eq.(\ref{densityratio}) to
\be
\frac{n'_\gamma(\epsilon_{\gamma_1})}
{n'_\gamma(\epsilon_{\gamma_2})}=\frac{\Phi_{SSC}(\epsilon_{\gamma
    1})}{\Phi_{SSC}(\epsilon_{\gamma 2})}
 \frac{E_{\gamma_1}}{E_{\gamma_2}}.
\label{denratio}
\ee
The $\gamma$-ray flux from the $\pi^0$ decay is deduced to be 
\be
F_{\gamma}(E_{\gamma}) \equiv E^2_{\gamma} \frac{dN(E_\gamma)}{dE_\gamma} 
\propto  E^2_p \frac{dN(E_p)}{dE_p} n'_{\gamma,f} .
\ee
The EBL effect attenuates the VHE flux by a factor of
$e^{-\tau_{\gamma\gamma}}$, where $\tau_{\gamma\gamma}$ is the optical
depth which depends on the energy of the propagating VHE $\gamma$-ray and the
redshift $z$ of the source.

Including the EBL effect, the relation between observed flux
$F_{\gamma}$ and the intrinsic flux $F_{int}$ is given as
\be
F_{\gamma}(E_{\gamma}) = F_{int}(E_{\gamma})
e^{-\tau_{\gamma\gamma}(E_{\gamma},z)}.
\label{fluxrelat}
\ee
Then the EBL corrected observed multi-TeV photon flux from $\pi^0$-decay
at two different observed photon energies $E_{\gamma 1}$ and
$E_{\gamma 2}$ can be expressed as
\be
\frac{F_\gamma(E_{\gamma_1})}{F_\gamma(E_{\gamma_2})} 
=
\frac{\Phi_{SSC}(\epsilon_{\gamma_1})}{\Phi_{SSC}(\epsilon_{\gamma_2})}
\left(\frac{E_{\gamma_1}}{E_{\gamma_2}}\right)^{-\alpha+3}
e^{-\tau_{\gamma\gamma}(E_{\gamma_1},z)+\tau_{\gamma\gamma}(E_{\gamma_2},z)},
\label{sscspectrum}
\ee
where we have used 
\be
\frac{E_{p_1}}{E_{p_2}}=\frac{E_{\gamma_1}}{E_{\gamma_2}}.
\ee
The $\Phi_{SSC}$ at different energies are calculated
using the leptonic model. Here the multi-TeV flux is proportional to
$E_{\gamma}^{-\alpha+3}$  and $\Phi_{SSC}(\epsilon_{\gamma})$. 
In the photohadronic process ($p\gamma$), the
multi-TeV photon flux is expressed as
\be
F(E_{\gamma})=A_{\gamma} \Phi_{SSC}(\epsilon_{\gamma} )\left (
  \frac{E_{\gamma}}{TeV}  \right )^{-\alpha+3}
e^{-\tau_{\gamma\gamma}(E_{\gamma},z)}.
\label{modifiedsed}
\ee
Both $\epsilon_{\gamma}$ and $E_{\gamma}$ satisfy the condition given in
Eq.(\ref{Eegamma}) and the dimensionless constant 
$A_{\gamma}$ is given by
\be
A_{\gamma}=\left( \frac{F(E_{\gamma_2})}{\Phi_{SSC}(\epsilon_{\gamma2}
    )}\right ) \left (
  \frac{TeV}{E_{\gamma_2}}  \right )^{-\alpha+3}e^{\tau_{\gamma\gamma}(E_{\gamma_2},z)}.
\label{agamma}
\ee
Comparing Eqs. (\ref{fluxrelat}) and (\ref{modifiedsed}), the intrinsic
flux $F_{int}$ is given as
\be
F_{int}(E_{\gamma})=A_{\gamma} \Phi_{SSC}(\epsilon_{\gamma} )\left (
  \frac{E_{\gamma}}{TeV}  \right )^{-\alpha+3}.
\label{fint}
\ee
Using Eq. (\ref{modifiedsed}), we can calculate the EBL corrected
multi-TeV flux where $A_{\gamma}$ can be fixed from observed flare
data. We can calculate the Fermi accelerated high energy proton flux $F_p$
from the TeV $\gamma$-ray flux through the relation\cite{Sahu:2012wv}
\be
F_p(E_p)=7.5\times \frac{F_{\gamma}(E_{\gamma})}{\tau_{p\gamma}(E_p)}.
\ee
The optical depth $\tau_{p\gamma}$ is given in Eq.(\ref{optdepth}).
For the observed highest energy $\gamma$-ray, $E_{\gamma}$ corresponding
to a proton energy $E_p$, the proton flux
$F_p(E_p)$ will be always smaller than the Eddington flux
$F_{Edd}$. This condition puts a lower limit on the optical depth of
the process and is given by
\be
\tau_{p\gamma} (E_p) > 7.5\times
\frac{F_{\gamma}(E_{\gamma})}{F_{Edd}}.
\label{optdepthvhe}
\ee
From the comparison of different times scales and from
Eq.(\ref{optdepthvhe}) we will be able to constrain the seed photon
density in the inner jet region.

It is observed that, for the observed
flare energy $E_{\gamma}$  the range of $\epsilon_{\gamma}$ 
is always in the low
energy tail region of the SSC band and the corresponding SSC flux in this range of seed
photon energy is exactly a
power-law given by $\Phi_{SSC} \propto \epsilon^{\beta}_{\gamma}$
with $\beta > 0$. Again, from the kinematical condition to produce
$\Delta$-resonance through $p\gamma$ interaction, $\epsilon_{\gamma}$
can be expressed in terms of $E_{\gamma}$ and can be written as
\be
\Phi_{SSC}(\epsilon_{\gamma}) =\Phi_0\, E^{-\beta}_{\gamma}.
\label{phi_power_law}
\ee
From the leptonic model fit to the observed multiwavelength data (up
to second peak) during a quiescent/flaring state we can get the SED
for the SSC region from which $\Phi_0$ and $\beta$ can be obtained easily. 
By expressing the observed flux $F_{\gamma}$ in terms of the
intrinsic flux $F_{\gamma,in}$ and the EBL correction as
\be
F_{\gamma} (E_{\gamma}) = F_{\gamma, in}(E_{\gamma})\,e^{-\tau_{\gamma\gamma}(E_{\gamma},z)}, 
\ee
where the intrinsic flux is given in ref. \cite{Sahu:2016bdu} as,
\be
F_{\gamma, in}(E_{\gamma}) =A_{\gamma}\,\Phi_0 \left (\frac{E_{\gamma}}{TeV}
\right )^{-\alpha-\beta+3},
\ee
where $A_{\gamma}$ is a dimensionless normalization constant and can be fixed
by fitting the observed VHE data. As discussed above the power
index $\beta$ is fixed from the tail region of the SSC SED for a given
leptonic model which fits the low energy data well. So the Fermi
accelerated proton spectral index $\alpha$ is
the only free parameter to fit the intrinsic spectrum.

\section{Results}

From the continuous monitoring and dedicated multi-wavelength
observations of the nearest HBLs Markarian 421 (Mrk 421, z=0.0308\cite{Gorham:1999en,Ulrich:1975ApJ}), Mrk 501 (z=0.033 \cite{Ulrich:1975ApJ,Stickel:1993aap}) and 1ES 1959+650 (z=0.047\cite{Schanchter:1993ApJ}), several major multi-TeV flares have been observed\cite{Tluczykont:2011gs,Aharonian:2000nr,Aliu:2016kzx,Chandra:2017vkw,Santander:2017wjl}. 
Strong temporal correlation in different wavebands, particularly in
X-rays and VHE $\gamma$-rays  has been observed in some flaring
events, however, in some other flaring events no such correlation is
observed \cite{Krawczynski:2003fq,Blazejowski:2005ih}, which seems unusual for a leptonic origin\cite{Katarzynski:2006db,Fossati:1998zn,Ghisellini:1998it,Roustazadeh:2011zz} of the multi-TeV
emissions and needs to be addressed through other alternative
mechanisms\cite{Mucke:1998mk,Mucke:2000rn,Cao:2014nia,Zdziarski:2015rsa,Essey:2010er,Ghisellini:2004ec,Tavecchio:2008be}. 
Below we shall discuss the flaring of Mrk 421, Mrk 501 and 1ES1959+650 separately.

\subsection{Markarian 421}

Mrk 421 was the first extragalactic source 
detected in the 
multi-TeV domain\cite{Punch:1992xw} and also it is one of the fastest varying $\gamma$-ray
sources. It has a luminosity distance $d_L$
of about 129.8 Mpc and its central supermassive black
hole is assumed to have  a mass $M_{BH}\simeq (2-9)\times 10^8\, M_{\odot}$
corresponding to a Schwarzschild radius of $(0.6-2.7)\times 10^{14}$
cm and the Eddington luminosity $L_{Edd}=(2.5-11.3)\times 10^{46}\, erg\,s^{-1}$. 
The synchrotron peak (1st peak) of its SED is in the soft to medium X-ray range and the
SSC peak (2nd peak) is in the GeV range. Through dedicated multi wavelength
observations, the source has been studied intensively. These studies
show a correlation between X-rays and VHE
$\gamma$-rays. A one-zone SSC model explains the observed SED
reasonably well\cite{Abdo:2011zz}. 
Several large flares were observed in 2000 -
2001\cite{Okumura:2002jw,Amenomori:2003dy,Fossati:2007sj} and 
2003 - 2004\cite{Cui:2004wi,Blazejowski:2005ih}.
During April 2004, a large flare took place both in the X-rays and the
TeV energy bands. The flare lasted for more than two weeks (from MJD 53,104
to roughly MJD 53,120). Due to a large data gap between MJD 53,093
and 53,104, it is difficult to exactly quantify the duration. The source was observed
simultaneously at TeV energies with the Whipple 10 m telescope and at
X-ray energies with the Rossi X-ray Timing Explorer (RXTE)\cite{Blazejowski:2005ih}. It was
also observed simultaneously at lower wavelengths (both radio and optical). 
During the flaring it was observed that, the TeV flares had no
coincident counterparts at longer wavelengths. Also
it was observed that the X-ray flux reached its peak 1.5 days before the
TeV flux did during this outburst. 
%Generally it is believed that, the TeV flare
%might not be a true orphan flare, like the one observed in 1ES
%1959+650.  
Remarkable similarities between
the orphan TeV flare in 1ES 1959+650 and Mrk 421 were observed,
including similar variation patterns in the X-ray spectrum.
A strong outburst in multi-TeV
energy in Mrk 421 was first detected by VERITAS telescopes on 16th of February
2010 and follow up observations were done by the HESS telescopes
during four subsequent nights\cite{Tluczykont:2011gs}.

A six month long multi-instrument campaign by the
MAGIC telescopes observed VHE flaring from Mrk 421 on 25th of
April 2014 and the flux (above 300 GeV) was about 16 times brighter
than the usual one. This triggered a joint ToO program by XMM-Newton,
VERITAS, and MAGIC instruments. These three instruments individually observed
approximately 3 h each day on April 29, May 1, and May
3 of 2014\cite{Abeysekara:2016qwu}. The simultaneous VERITAS-XMM-Newton observation is
published recently and it is shown that the observed multiwavelength spectra are
consistent with one-zone synchrotron self-Compton
model\cite{Abeysekara:2016qwu}. 
%However, 
%the details of the large flare observed on 25th April by MAGIC are not
%publicly available yet.

\begin{figure}%fig2
%\vspace{-0.3cm}
{\centering
\resizebox*{0.8\textwidth}{0.5\textheight}
%\resizebox*{0.8\textwidth}{0.5\textheight}
{\includegraphics{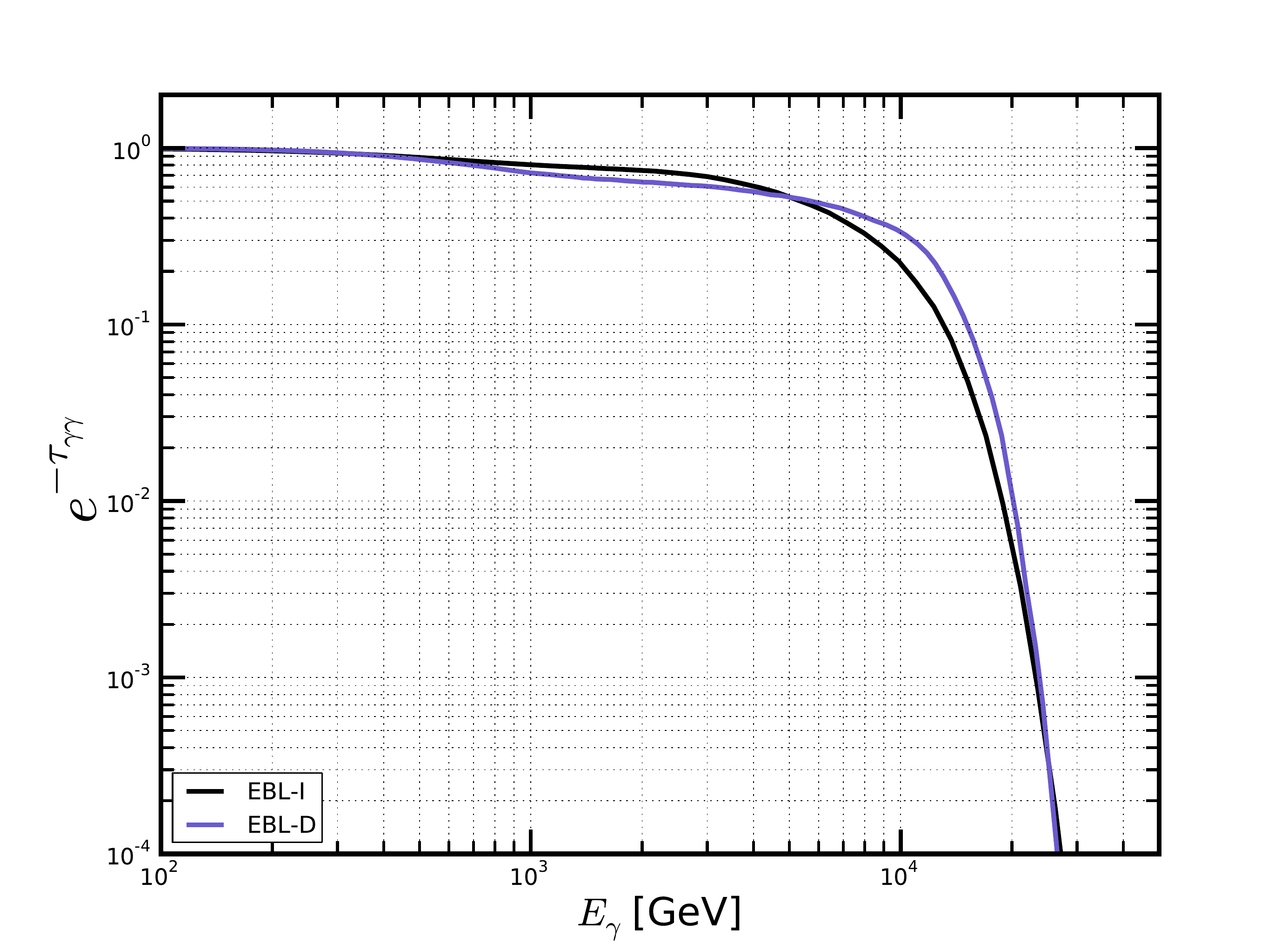}}
\par}
\caption{ 
%The attenuation factor as a function of photon energy predicted by the two EBL models for $z=0.031$.
At a redshift of $z=0.031$, the attenuation factor
as a function of 
$E_{\gamma}$ for different EBL models are shown for comparison. 
\label{fig:figure1}
}
\end{figure}

The observed multi-TeV flux from the above flaring events are
EBL corrected and for this correction well known EBL models of Dominguez et
al.\cite{Dominguez:2010bv} (EBL-D) and Inoue et al.\cite{Inoue:2012bk}
(EBL-I) are used. In Fig. \ref{fig:figure1} the attenuation factor
for these two models as functions of observed gamma-ray energy
$E_{\gamma}$ are shown and both are
practically the same in all the energy ranges (there is
a minor difference in the energy range $600\, {\text GeV} \leq
E_{\gamma} \leq 1\, {\text TeV}$). 
%%%%%%%%%%%%%%%%%%%%%%%%%%%%%%%%%%%%%%%%%%%
%%%%%%%%%%%%%%%%%%%%%%%%%%%%%%%%%%%%%%%%%%%

\subsubsection{ The flare of April 2004}

\begin{figure}%fig2
%\vspace{-0.3cm}
{\centering
\resizebox*{0.8\textwidth}{0.5\textheight}
%\resizebox*{0.8\textwidth}{0.5\textheight}
{\includegraphics{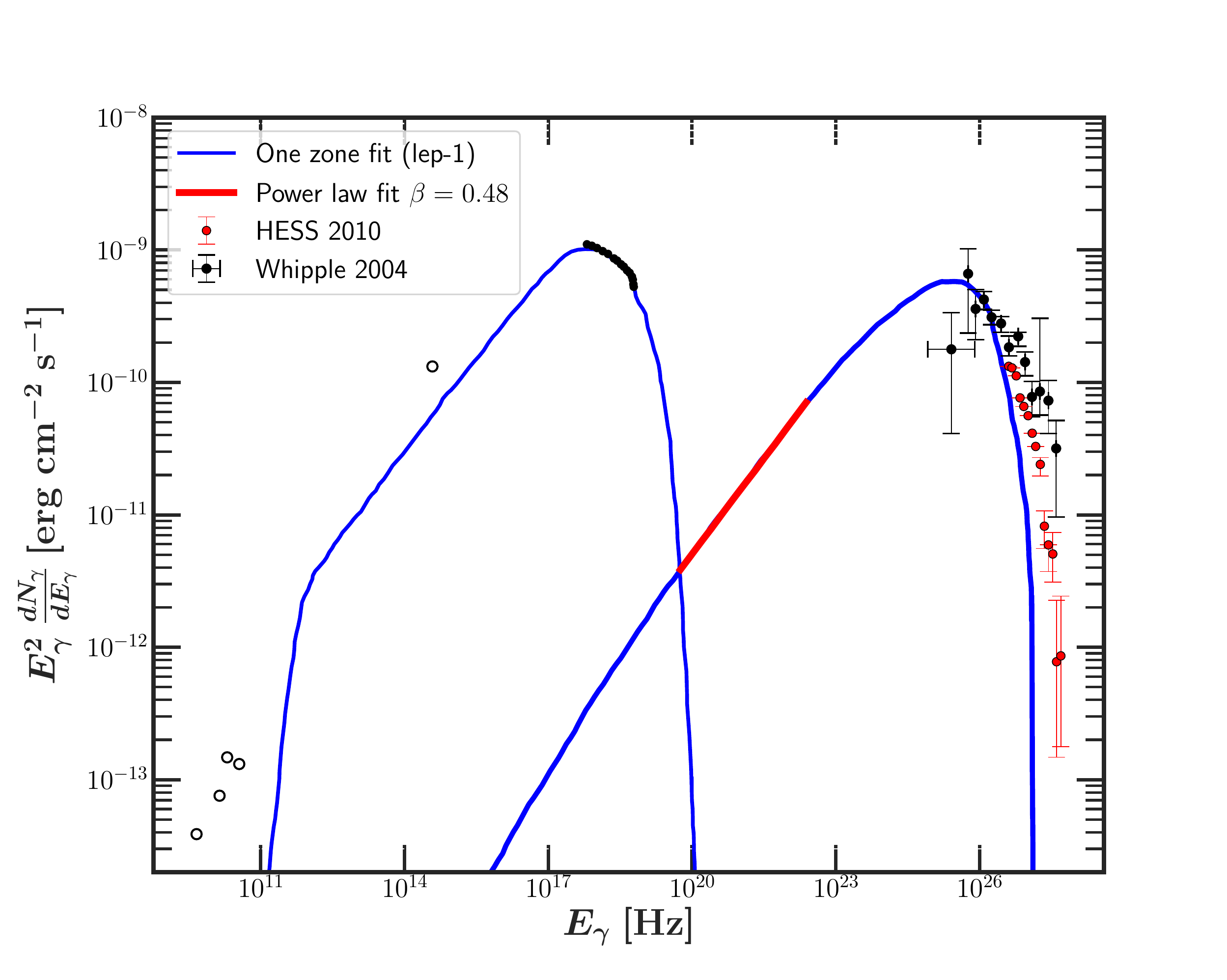}}
\par}
\caption{
%The observed data to the extreme left (from radio to infrared frequency
%  range, open circles)  are from quiescent state observed by the University of
%  Michigan radio astronomy observatory (UMRAO) during 2003-2004.
The fit to the tail region of the SSC SED (lep-1)\cite{Blazejowski:2005ih} with the power-law
as given in Eq. (\ref{phi_power_law}) with $\Phi_0=6.0\times
10^{-10}\, erg\, cm^{-2}\, s^{-1}$ and $\beta=0.48$ (red curve).
\label{fig:figure2}
}
\end{figure}

The multi-TeV flare of April 2004 was the first flare observed in
multiwavelength by the Whipple telescope in the energy range $ 0.25\,
{\rm TeV} (6.0\times 10^{25}\, {\rm Hz}) \le E_{\gamma} \le 16.85\,
{\rm TeV}  (4.1\times 10^{27}\, {\rm Hz})$ and it was difficult to explain by one-zone
leptonic model\cite{Blazejowski:2005ih}. As discussed above, the
photohadronic interpretation of the flare data needs the leptonic
model as input and here the one-zone leptonic model 
of ref. \cite{Blazejowski:2005ih} (lep-1) is used. In this model the bulk Lorentz is
$\Gamma={\cal D}=14$. The above range of $E_{\gamma}$
corresponds to the Fermi accelerated proton energy  in the range $2.5\, {\rm TeV} \le
E_p \le 168\, {\rm TeV}$ and the corresponding background photon energy is  
in the range $23.6\, {\rm MeV} (5.7\times 10^{21}\, {\rm Hz})$ $\ge$
 $\epsilon_{\gamma} \ge 0.35\, {\rm MeV}$ $ (8.4\times 10^{19}\, {\rm Hz}) $. This
range of $\epsilon_{\gamma}$ is in the low energy tail 
region of the SSC SED and its flux is expressed as
power-law given in Eq. (\ref{phi_power_law}) 
with $\Phi_0 = 6.0\times 10^{-10}$ $ {\rm erg\, cm^{-2}\, s^{-1} }$ and
$\beta=0.48$ which is shown in Fig. \ref{fig:figure2}. 

%%%%%%%%%%%%%%%%%%%%%%%%%%%%%%%%%%%%%%%%%%%
%%%%%%%%%%%%%%%%%%%%%%%%%%%%%%%%%%%%%%%%%%%

%%%%%%%%%%%%%%%%%%%%%%%%%%%%%%%%%%%%%%%%%%%
\begin{figure}%fig3
%\vspace{-0.3cm}
{\centering
\resizebox*{0.8\textwidth}{0.5\textheight}
%\resizebox*{0.8\textwidth}{0.5\textheight}
{\includegraphics{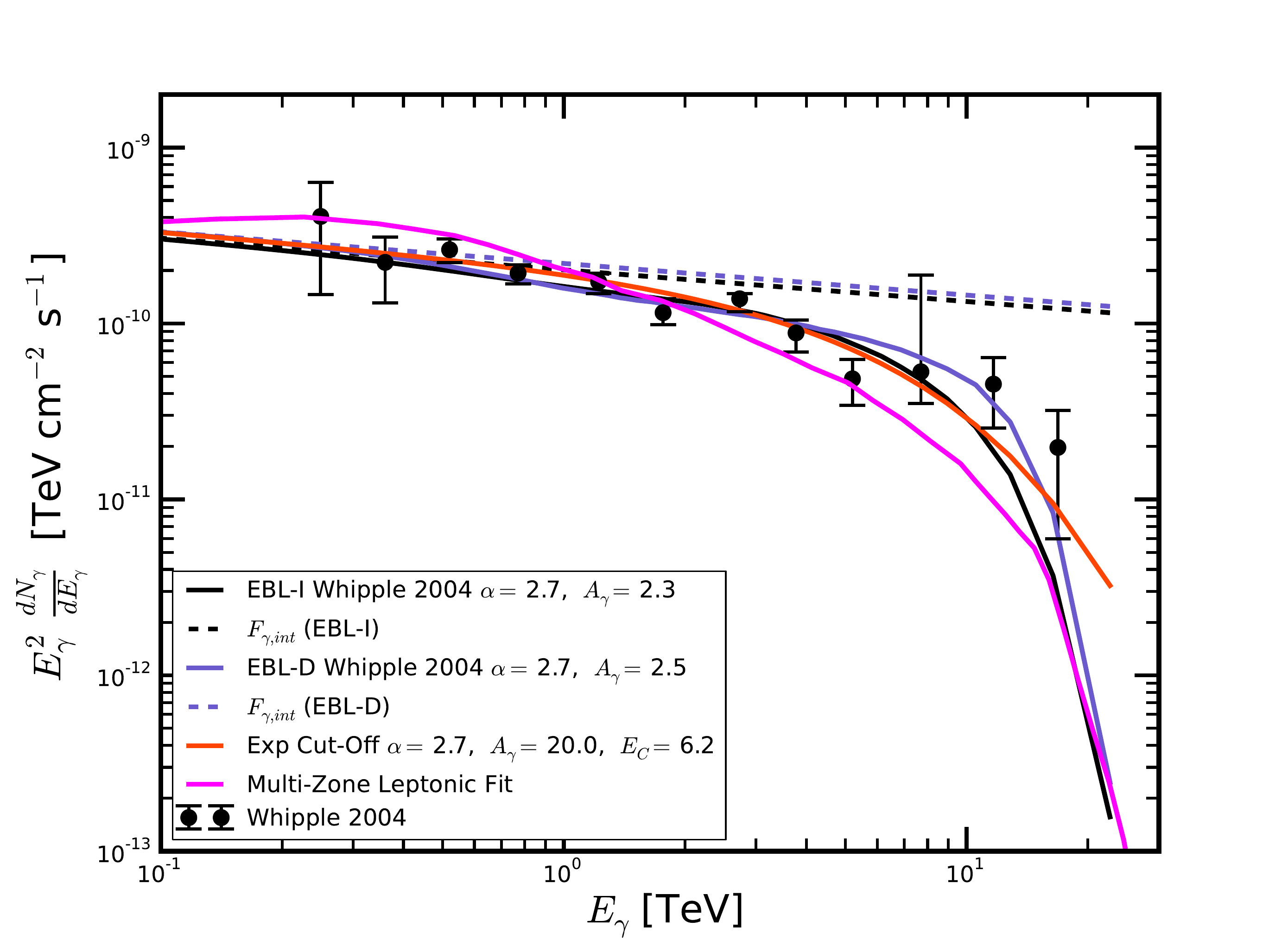}}
\par}
\caption{
Fit to the observed flux of April 2004 flare  with the photohadronic model
using two different EBL models are shown. It is also compared with the power-law
with exponential cut-off without EBL correction fit\cite{Sahu:2015tua} and with the
multi-zone leptonic fit\cite{Blazejowski:2005ih}. The multi-zone
leptonic model accounts for the attenuation of the very high energy 
gamma-rays by  the diffuse infrared background.The intrinsic fluxes for
both the EBL models are also shown.
\label{fig:figure3}
}
\end{figure}
%%%%%%%%%%%%%%%%%%%%%%%%%%%%%%%%%%%%%%%%%%%

Very good fit to the multi-TeV flare data is obtained for  
$\alpha=2.7$, and the
normalization constant $A_{\gamma}$ for EBL-D is  2.3 and for EBL-I
is 2.5 respectively which are shown in
Fig. \ref{fig:figure3}. Below 4 TeV
both these EBL models fit the data very well, and, above this energy 
there is a slight difference due to the change in the attenuation factor.
It is observed that EBL-D fits better than the EBL-I. To compare 
the photohadronic
model fit without EBL correction but with an exponential
cut-off ($dN/dE_{\gamma}\propto E_{\gamma}^{-\alpha} exp(-E_{\gamma}/E_c)$) \cite{Sahu:2015tua}, and the multi-zone leptonic model fit
(magenta curve)\cite{Blazejowski:2005ih} are also shown in the same
figure (red curve). It can be seen from Fig. \ref{fig:figure3} that the
multi-zone fit is not so good compared to other fits for $E_{\gamma}
\leq 15$ TeV. However, for higher energy it has the same behavior as EBL-D
and EBL-I. With the exponential cut-off scenario, 
a good fit is obtained for the spectral index $\alpha=2.7$ and the cut-off
energy $E_c=6.2$ TeV. Again comparing this with the EBL corrected models,
below 4 TeV,  all these
fits are exactly the same. However, above 4 TeV we observe some 
discrepancy among these fits and above 10 TeV the fits of EBL-D
and EBL-I fall faster than the exponential cut-off
scenario. Comparison of EBL-I with the
exponential cut-off  scenario shows that, for $E_{\gamma}\leq 10$ TeV both
$e^{-E_{\gamma}/E_c}$ and $e^{-\tau_{\gamma\gamma}}$  are almost the same and
above 10 TeV the attenuation factor falls faster than the
exponential cut-off. The intrinsic flux in EBL-D and EBL-I are
almost the same and having power-law behavior with $F_{\gamma,
  in}\propto E^{-0.18}_{\gamma}$. Even though  all these models fit
quite well to the observed data below 20 TeV energy range, the deviation is
appreciable above 20 TeV between the EBL corrected plots and the exponential
cut-off. So observation of VHE flux above $\sim 30$
TeV will be a good test to constrain the EBL effect on the VHE
gamma-rays from Mrk 421.

By comparing different time scales i.e. expansion time scale,
interaction time scale of $p\gamma$ interaction and using the fact
that the high energy
proton luminosity to be smaller than the Eddington luminosity in the
inner jet region of size $R'_f\simeq 3\times 10^{15}\, {\rm cm}$, the range of
optical depth for the $\Delta$-resonance production is estimated as 
$0.02 \, < \tau_{p\gamma} \, < 0.13$. This corresponds to a photon
density in the inner jet region as $1.3\times 10^{10}\, {\rm cm^{-3}} <
n'_{\gamma, f} < 8.9\times 10^{10}\, {\rm cm^{-3}}$. 
The TeV photons produced from the neutral pion decay will mostly
encounter the SSC photons in the energy range $0.35\, MeV \leq
\epsilon_{\gamma} \leq 23.6\, MeV$. The pair production cross section
for $\epsilon_{\gamma} \geq 0.35\, MeV$ is very small
($\sigma_{\gamma\gamma} \leq 10^{-30}\, cm^{-2}$) which corresponds to a mean free path of
$\lambda_{\gamma\gamma} \geq 10^{19}\, cm$ for the
multi-TeV gamma-rays, larger than the outer jet size. So, the TeV
photons will not be attenuated much due to the $e^+e^-$ pair
production. 

%The parameters used in
%the photohadronic to fit the 2004 data are summarized in Table \ref{tab1}.
%Also it is shown that
%TeV photons produced from the neutral pion decay will not be attenuated due to
%the low density of low energy (0.05-3.5 eV) photons in the production
%site. The above situation is also valid here. 

\subsubsection{The flare of February 2010}

On 16th of February 2010, 
a strong outburst in multi-TeV gamma-rays from Mrk 421 was observed by VERITAS
telescopes  and follow up observations were
carried out by HESS telescopes from 17th to 20th of February a total
of 6.5 h. These data were taken in 11 runs with each run $\sim 28$
minutes duration \cite{Tluczykont:2011gs}. The HESS telescopes
observed the flare in the energy range $1.67\, {\rm TeV} (4.0\times
10^{26}\, {\rm Hz}) \le E_{\gamma} \le 20.95\, {\rm TeV} (5.0\times
10^{27}\, {\rm Hz})$. During this period there was no observation of
multiwavelength SED, particularly in the SSC band. So to
interpret the observed flare data, the observed SED at an earlier and
later epochs were used. The first one is the lep-1, which is
used for the interpretation of the April 2004 flare and the second
leptonic SED is from the multiwavelength observation of Mrk 421 during 
January to March 2013, undertaken by GASP-WEBT, {\it Swift}, NuSTAR
Fermi-LAT, MAGIC, VERITAS\cite{Balokovic:2015dnz} and fitted with one-zone leptonic model
where the bulk Lorentz factor $\Gamma=25$ is used (lep-2). The
parameters of lep-1 and lep-2 are shown in Table-\ref{tab1}.
%%%%%%%%%%%%%%%%%%%%%%%%%%%%%%%%%%%%%%%%%%%
\begin{figure}%fig5
%\vspace{-0.3cm}
{\centering
\resizebox*{0.8\textwidth}{0.5\textheight}
%\resizebox*{0.8\textwidth}{0.5\textheight}
{\includegraphics{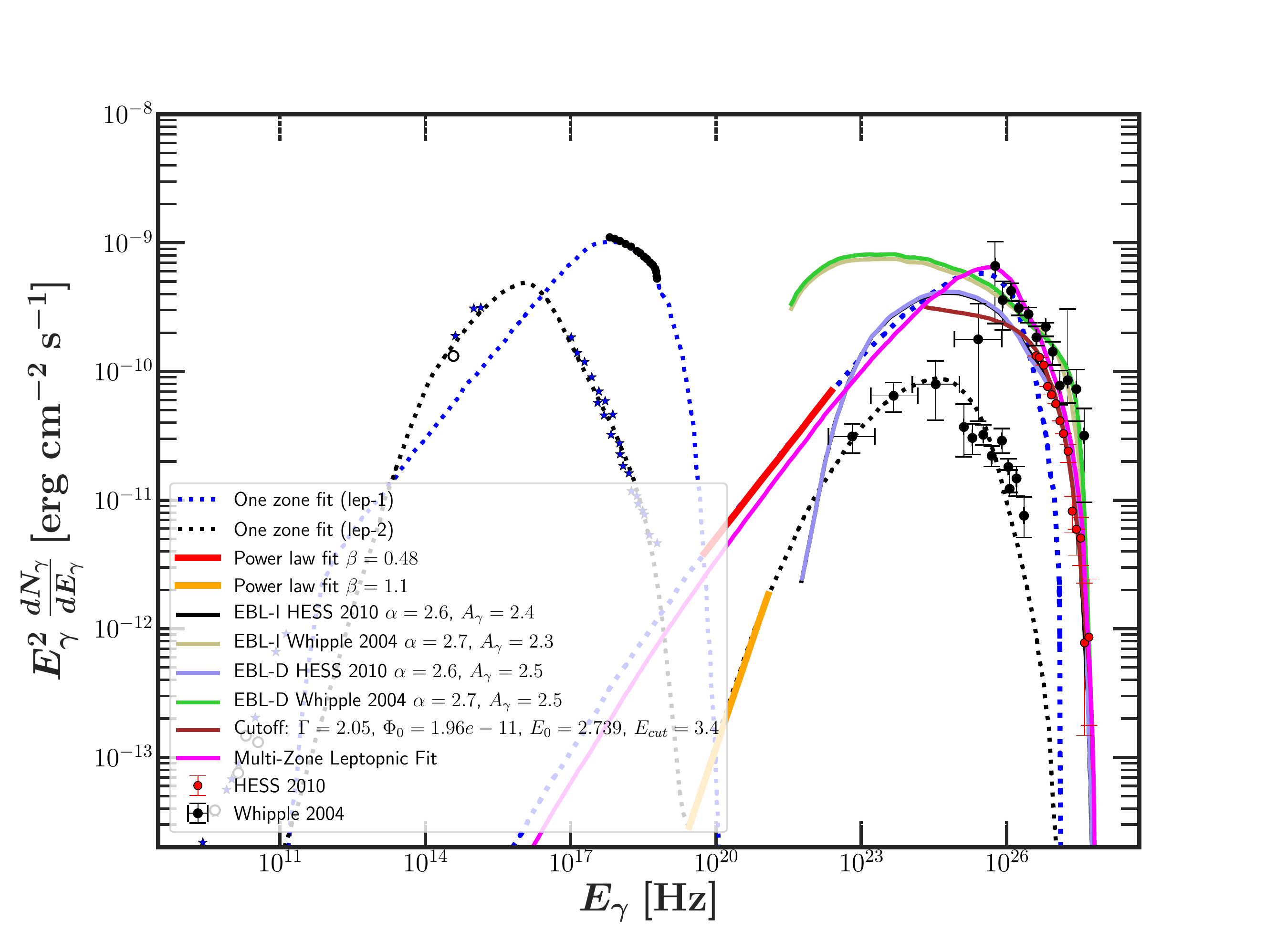}}
\par}
\caption{
The SED of lep-2\cite{Balokovic:2015dnz} is shown along with the power-law fit to the
  SSC tail region with $\beta=1.1$. The best fit to the flare
 of 2010 using EBL-D and EBL-I are also shown. For comparison, we have
 also shown the SED of lep-1, the power-law fit to the SSC tail region
 with $\beta=0.48$ and the best fit to the flare data of 2004 by
 Whipple telescope\cite{Blazejowski:2005ih}. The low energy observed
 data are taken from ref. \cite{Balokovic:2015dnz,Blazejowski:2005ih}.
\label{fig:figure5}
}
\end{figure}
%%%%%%%%%%%%%%%%%%%%%%%%%%%%%%%%%%%%%%%%%%%

%%%%%%%%%%%%%%%%%%%%%%%%%%%%%%%%%%%%%%%%%%%
\begin{figure}%fig6
%\vspace{-0.3cm}
{\centering
\resizebox*{0.8\textwidth}{0.5\textheight}
%\resizebox*{0.8\textwidth}{0.5\textheight}
{\includegraphics{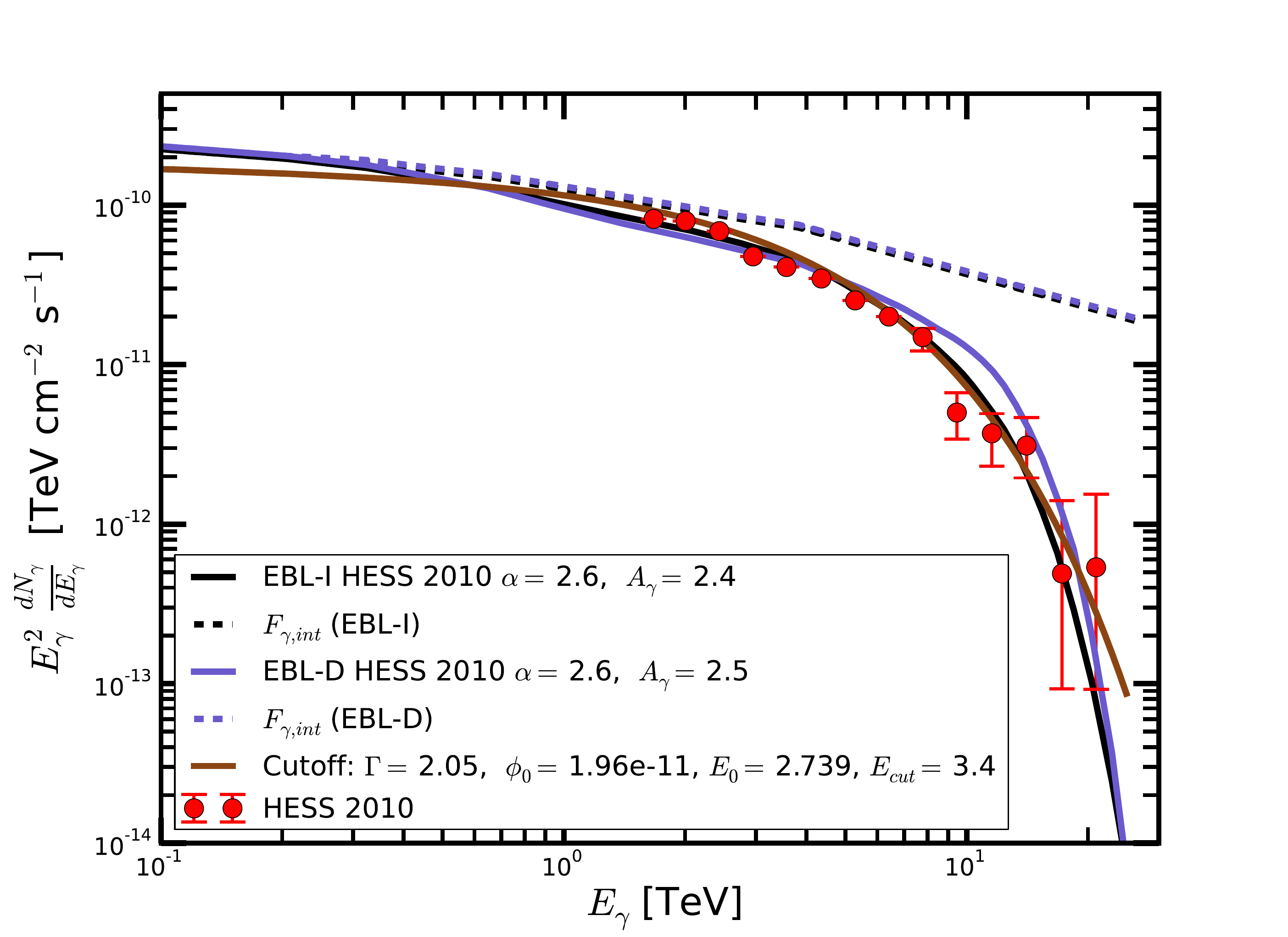}}
\par}
\caption{
Fit to the observed flux of 2010 flare by HESS  using photohadronic model
and EBL correction to it by EBL-D and EBL-I are shown. The corresponding intrinsic fluxes are
also given.
\label{fig:figure6}
}
\end{figure}
%%%%%%%%%%%%%%%%%%%%%%%%%%%%%%%%%%%%%%%%%%%
%%%%%%%%%%%%%%%%%%%%%%%%%%%%%%%%%%%%%%%%%%%

\begin{table}[h]
\centering
\caption{The parameters used in the photohadronic fits
  for the observed data taken from the observations of Whipple in 2004
  and HESS in 2010 are given. The parameters $\alpha$ and $A_{\gamma}$ are spectral index and
  normalization constant respectively.} 
\label{tab1}
\begin{tabular*}{\columnwidth}{@{\extracolsep{\fill}}llll@{}}
\hline
\multicolumn{1}{@{}l}{} & lep-1 & lep-2\\
\hline
${\cal D}$ (Doppler factor)& 14 & 25\\
$R^{\prime}_b$ (Blob radius) & $0.7 \times 10^{16} cm$ & $0.9 \times 10^{16} cm$& \\
$R^{\prime}_f$ (Inner blob radius)& $ \approx 3 \times 10^{15} cm$ & $ \approx 3 \times 10^{15} cm$\\
$B^{\prime}$ (Magnetic field)& 0.26 G& 0.17 G\\
\hline
EBL Model & $\alpha$,  $A_{\gamma}$ & $\alpha$,  $A_{\gamma}$ \\
\hline
EBL-I &2.7, 2.3 & 2.6, 2.4\\
EBL-D & 2.7, 2.5 & 2.6, 2.5 \\
\hline
\end{tabular*}
\end{table}

Again, the above range
of $E_{\gamma}$ corresponds to the  proton energy in the
range $16.7\, {\rm TeV} \le E_p \le 210\, {\rm TeV}$. Using lep-1, where
$\Gamma={\cal D}=14$, the seed
photon energy lies in the range $0.28\, {\rm MeV} (6.8\times 10^{19}\, {\rm Hz})
\le \epsilon_{\gamma} \le 3.53\, {\rm MeV} (8.5\times 10^{20}\, {\rm Hz})$ which
is again in the tail region of the SSC SED as shown in
Fig. \ref{fig:figure2}. Very good fit to the multi-TeV spectrum is obtained by using the
EBL-D and EBL-I and the parameters are respectively $\alpha=3.1$ and
$A_{\gamma}=58.0$ for EBL-D and $\alpha=3.2$ and
$A_{\gamma}=28.0$ for EBL-I which correspond to very soft spectrum and
the intrinsic spectrum is also soft (between -0.68 to -0.58). In the
low energy limit the spectrum shoots up very high and this behavior is
not seen by HESS. So we can ignore the fit for which $\alpha > 3$.
This soft power-law problem arises
because $\beta=0.48$ is small and we can use the leptonic models which
have $\beta > 0.48$ as a result of which we should get $\alpha < 3$.
The time averaged differential energy spectrum of this 
observation is also fitted with  a power-law with exponential cut-off
having four parameters\cite{Tluczykont:2011gs}.

For lep-2 with $\Gamma={\cal D}=25$, the energy range 
$1.67\, {\rm TeV} (4.0\times 10^{26}\, {\rm Hz}) \le E_{\gamma} \le
20.95\, {\rm TeV} (5.0\times 10^{27}\, {\rm Hz})$ corresponds to the
seed photon energy in the range 
$0.90\, {\rm MeV} (2.17\times 10^{20}\, {\rm Hz})$ $
\le \epsilon_{\gamma} \le 11.26\, {\rm MeV} (2.72\times 10^{21}\, {\rm Hz})$,
which is again in the tail region of the SSC SED as shown in
Fig. \ref{fig:figure5}. This is fitted with a 
power-law with $\beta=1.1$ and $\Phi_0=4.37\times 10^{-9}\, {\rm TeV} \,
{\rm cm^{-2}\, s^{-1}}$. 
Again EBL-D and EBL-I are used to fit the
2010 flare data in the photohadronic model which are also shown in
Fig. \ref{fig:figure5}. The best fit parameters are $\alpha=2.6$ and
$A_{\gamma}=2.5$ for EBL-D and $\alpha=2.6$ and
$A_{\gamma}=2.4$ for EBL-I respectively also the flux decreases towards
low energy regime as expected. 
%with a peak flux of $F_{\gamma,peak}\sim 2.6\times
%10^{-10}\, {\rm TeV} \, {\rm cm^{-2}\, s^{-1}}$ at 
%$E_{\gamma}\sim 18\, {\rm GeV}$. 
Both the EBL-D and EBL-I corrections to the
photohadronic model give practically the same result.

In the low energy
regime, the lep-1 fit flux increases drastically but this behavior 
is absent with the lep-2 fit.
%Comparison of the fits using lep-1 and lep-2 clearly shows that,
%only fitting to multi-TeV data is not enough. The fit has to be
%consistent in the low energy regime also which is absent in the
%lep-1 scenario. 
It is to be noted that, lep-1 corresponds to the
observation during the year 2003-2004 and lep-2 is the recent one
of January 2013. So we believe that during each observation period,
the photon density distribution in the jet changes and this change in
the seed photon changes the spectral behavior of the observed
multi-TeV gamma-rays. This implies almost simultaneous observation
in multiwavelength is essential to fit the observed data.
A minor difference between EBL-D and EBL-I
predictions for $0.6\, {\rm TeV} \leq E_{\gamma} \le 20\, {\rm TeV}$
is observed for 2010 flare data as shown in Fig. \ref{fig:figure6}
%This is due to the difference in attenuation factor in both these models.
and the intrinsic flux is $F_{\gamma,in}\propto
E^{-0.7}_{\gamma}$. Comparison of the intrinsic flux
$F_{\gamma,in}$ of 2004 and 2010 multi-TeV flaring shows that
different spectral shape of the observed events are solely due to the
diversity in the shape of the seed photon density distribution (particularly in the SSC
tail region) during different epochs.

\subsection{Markarian 501}

Markarian 501 (RA:$251.46^{\circ}$, DEC:$39.76^{\circ}$) is 
one of the brightest extragalactic sources in
X-ray/TeV sky\cite{Abdop:2010} and also 
the second extragalactic object (after
Mrk 421) identified as VHE emitter by
Whipple telescope in 1996. Since its discovery, the multiwavelength
correlation of Mrk 501 has been studied intensively and during this
period it  has undergone many major outbursts on long time scales and
rapid flares on short times scales mostly in the X-rays and TeV
energies\cite{Pian:1998hh,Krawczynski:1999vz,Tavecchio:2001,Ghisellini:2002ex,Sambruna:2000ic,Gliozzi:2006qq,Villata:1999,Katarzynski:2001,Aharonian:1998jw,Aharonian:1999vy,Aharonian:2000xr}. 
%It has been observed that, during these outbursts, both
%peaks have shifted to higher energies and during the most extreme case
%the synchrotron peak $\sim$ keV range has shifted above 200 keV\cite{Acciari:2010aa}. Due
%to the low sensitivity of the previous generation instruments, Mrk 501
%was primarily observed in VHE band during the outbursts. However,
%later on it was observed in all the wave bands. 
In the year 2009, Mrk 501 was observed as a
part of large scale multiwavelength campaign covering a period of 4.5
months (from March 9 to August 1, 2009) \cite{Aliu:2016kzx}.
The scientific goal of this extended observation was to collect a
simultaneous, complete multifrequency data set to test the
current theoretical models of broadband blazar emission mechanism. 
%Also this will help to understand the origin of high energy emission
%from blazars and the physical mechanism responsible for the
%acceleration of the charged particles in the relativistic jets. 
Between April 17 to May 5, this HBL was observed by  both 
space and ground based observatories, covering the entire
electromagnetic spectrum even including the variation in optical
polarization\cite{Aliu:2016kzx}. A very strong VHE flare was detected on May 1st first by Whipple
telescope and 1.5 hours later with VERITAS. 
Both of these
telescopes continued simultaneous observation of this VHE flare until
the end of the night and the detected flux was
enhanced by a factor of $\sim10$ above the average baseline flux.
Also a dramatic increase in
the flux by a factor $\sim 4$ in 25 minutes and a falling time of $\sim
50$ minutes was observed. The flux measured at lower energies before
and after the VHE flare did not show any significant 
variation. But, {\it Swift}-XRT (in X-ray) and UVOT (in optical) 
did observe moderate flux variability\cite{Aliu:2016kzx}. 
%Also both Whipple and VERITAS
%did observe statistically significant variation in VHE band.
Using the one-zone SSC model, the average
SED of this multiwavelength (up to second peak) campaign of Mrk 501 is interpreted satisfactorily\cite{Aliu:2016kzx}. 

The very strong VHE flare data of May 1st observed by Whipple
telescope  and the long outburst observed by HEGRA telescopes
in 1997 were modeled using the photohadronic model of
ref.\cite{Sahu:2016mww}. The model of Dominguez et
al.\cite{Dominguez:2010bv} is used to correct for the EBL effect on
the observed data. Here we shall discuss about the fit to this flare data
by different models.
%Our aim here is to use the photohadronic model  of Sahu et al.\cite{Sahu:2013ixa,Sahu:2012wv,Sahu:2013cja,Sahu:2015tua,Sahu:%2016bdu} and the EBL model of
%Dominguez et al.\cite{Dominguez:2010bv} to interpret the observed We found that both these flares can be well explained
%with this model.
To explain the VHE flare of May 1st, 2009, we use the
parameters of the one-zone leptonic model\cite{Aliu:2016kzx} whose
parameters are shown in Table \ref{tab2}.
%%%%%%%%%%%%%%%%%%%%%%%%%%%%%%%%
\begin{table}
\centering
\caption{The parameters (up to $B'$) are taken from the one-zone synchrotron
model of ref. \cite{Aliu:2016kzx} which are used to fit the SED of
Mrk 501. The last two parameters are obtained from the best
fit to the observed Whipple high state flare data  from ref.\cite{Sahu:2016mww}.
} 
\label{tab2}
\begin{tabular*}{\columnwidth}{@{\extracolsep{\fill}}llll@{}}
\hline
\multicolumn{1}{@{}l}{Parameter} &Description & Value\\
\hline
$M_{BH}$ & Black hole mass & $(0.9-3.5)\times 10^9 M_{\odot}$\\
z & Redshift & 0.034\\
%$\theta_{ob}$ &Viewing angle  &$10^{\circ}$\\
$\Gamma$ &Bulk Lorentz Factor & 12\\
${\cal D}$& Doppler Factor & 12\\
$R^{\prime}_b$ & Blob Radius & $1.2\times 10^{16}$cm\\ 
$B^{\prime}$ &Magnetic Field & $0.03$ G\\ 
\hline
$R'_f$ &Inner blob Radius& $5\times 10^{15}$cm\\
$\alpha$ &Spectral index& $2.4$\\
%$E_c$ &$\gamma$-ray Cut-off Energy& $30$ TeV\\
\hline
\end{tabular*}
\end{table}
%%%%%%%%%%%%%%%%%%%%%%%%%%%%%%%%%%%%%%%%%%%%%%%

%%%%%%%%%%%%%%%%%%%%%%%%%%%%%%%%%%%%%%%%%%%
%%%%%%%%%%%%%%%%%%%%%%%
\begin{figure}[t!]%fig2
%\vspace*{-0.1cm}
{\centering
\resizebox*{1.0\textwidth}{0.72\textheight}
%{\includegraphics[angle=-90]{Mrk501-TEVSED-re-HEGRA.eps}}
{\includegraphics{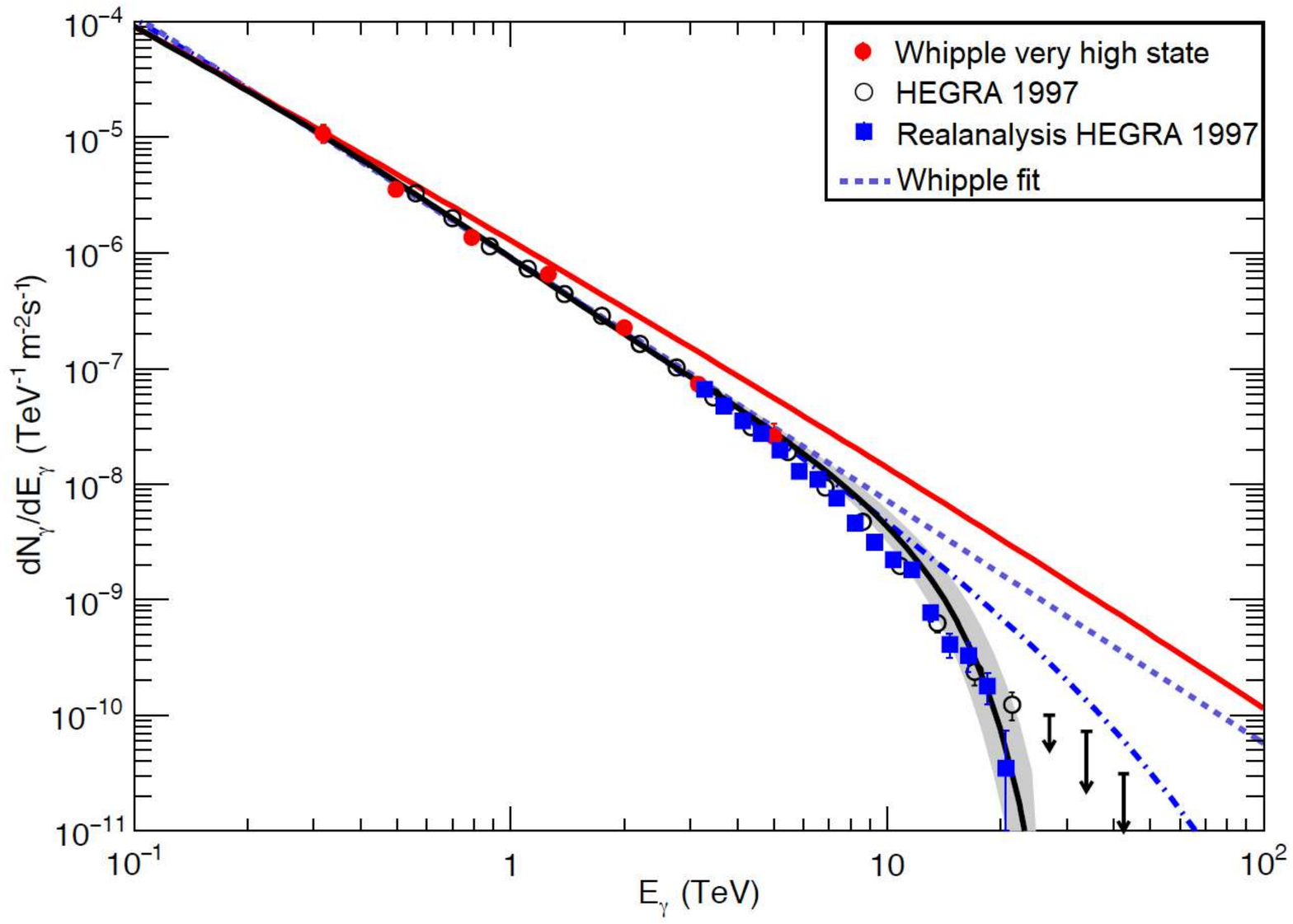}}
\par}
\vspace*{-0.50cm}
\caption{The black curve is the hadronic model fit which includes the
  EBL attenuation using the EBL model of Dominguez et al.\cite{Dominguez:2010bv} to the
  Whipple very high state flare data (red filled circles) of Mrk 501
  and the red continuous curve is the intrinsic flux in the same
  model. For comparison we have also shown the Whipple fit to the data
  (dashed curve) and the exponential fit (dashed dotted curve). 
We have also shown the HEGRA observation of the outburst during 1997:
conventional analysis (open circles)\cite{Aharonian:1999vy} and new analysis with improved
energy resolution (blue filled squares)\cite{Aharonian:2000xr}.
The shedded region is the region of uncertainty in the EBL model.
%(See the electronic edition of the Journal for a color version of this figure).
}
\label{mrk501tevsed}
\end{figure}
%%%%%%%%%%%%%%%%%%%%%%%%%%%%%%%%%%%%%%%%%%%
%%%%%%%%%%%%%%%%%%%%%%%%%%%%%%%%%%%%%%%%%%%
%%%%%%%%%%%%%%%%%%%%%%%
\begin{figure}[t!]%fig1
%\vspace*{-0.1cm}
{\centering
\resizebox*{1.0\textwidth}{0.65\textheight}
%{\includegraphics[angle=-90]{Mrk421-SED.eps}} % This is for rotation
{\includegraphics{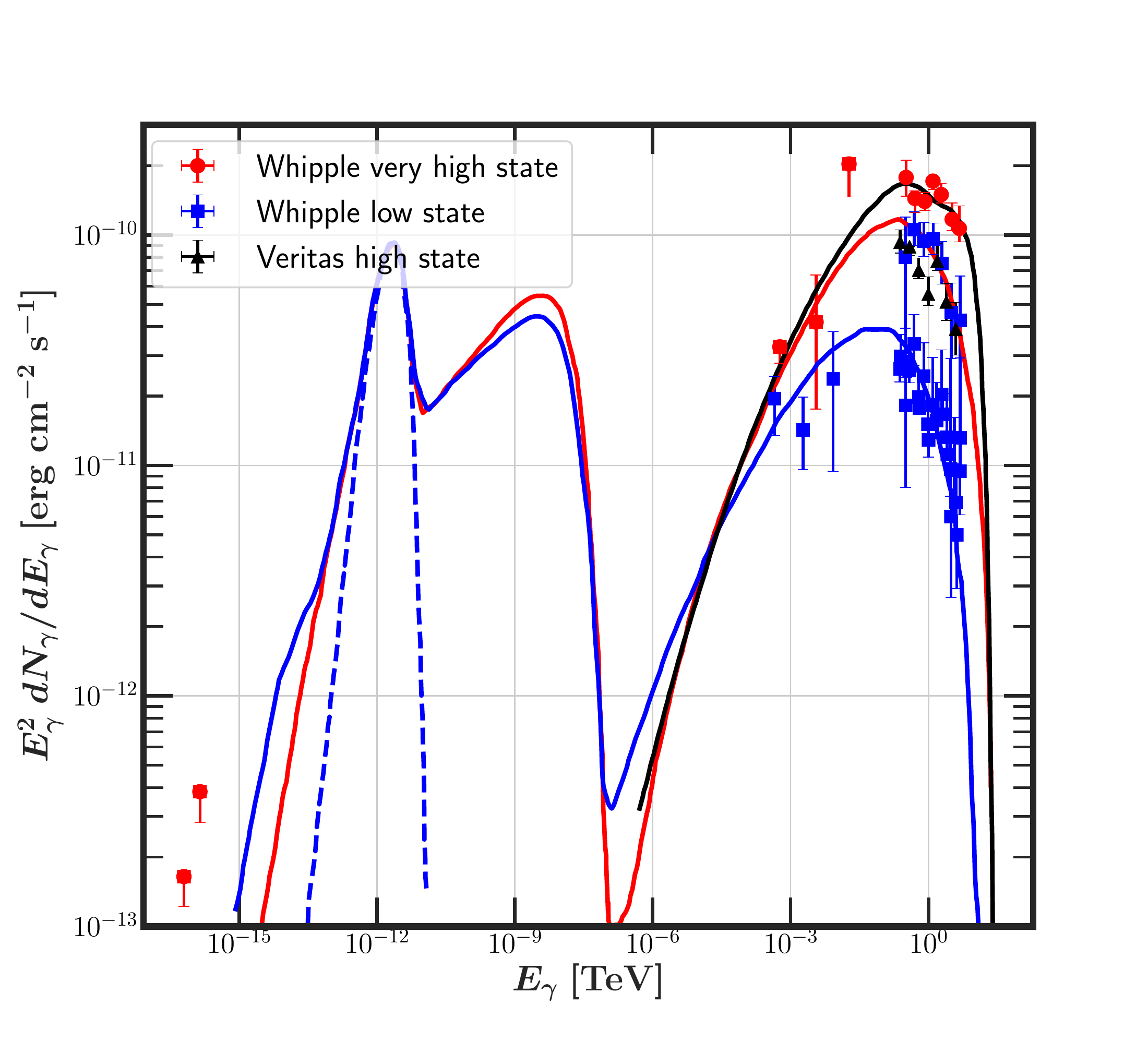}}
\par}
\vspace*{-0.30cm}
\caption{
The average SED of Mrk 501 is shown in all the energy bands which are
  taken from Ref. \cite{Aliu:2016kzx}. The SED of low state (MJD
  54936-54951; blue squares) and high state (MJD 54952-55; red
  circles) of the 3-week period are shown. The leptonic model fit to
  the low state (blue curve) and high state (red
  curve) are also shown. The blue dotted curve corresponds to the
  optical emission from the host galaxy. The black curve is the
  photohadronic fit to the Whipple very high state data (red circles). 
}
\label{mrk501sed}
\end{figure}
%%%%%%%%%%%%%%%%%%%%%%%%%%%
The observed VHE flare of May 1st
was in the range $\sim 317\,GeV \le E_{\gamma}\le\,5 \,TeV$. In the
present context, the above range of $E_{\gamma}$ corresponds to the
Fermi accelerated proton energy  in the range $3.2\, TeV\le
E_p \le 50\, TeV$ which interacts with
the inner jet background SSC photons in the energy range 
$13.6\, MeV (3.29\times 10^{21}\, Hz)\geq \epsilon_{\gamma} \geq\,\,
0.86 \, MeV  (2.1\times 10^{20}\, Hz)$ and finally produce the
observed VHE photons.
The range of $\epsilon_{\gamma}$ lies in the beginning of
the SSC spectrum. A very good fit to the
data is obtained for $\alpha=2.4$ and
$A_{\gamma}=89$ in Eq. (\ref{modifiedsed}). 
For comparison we have fitted the
data with an exponential cut-off function (dashed dotted curve) and the best fit is obtained for
$\alpha=2.6$, $E_c=30$ TeV  and $A_{\gamma}=66$. Also we have shown the 
Whipple fit (dashed curve) for comparison, where it is fitted by the
function $dN_{\gamma}/dE_{\gamma}=9.1\times 10^{-7} (E_{\gamma}/1 TeV)^{-2.1}\,
\text {ph}\, \text {m}^{-2}\, \text {s}^{-1}\, \text {TeV}^{-1}$. It is observed
that the EBL correction to the VHE
$\gamma$-ray is small but not insignificant (black curve in
Fig. \ref{mrk501tevsed}) 
which has a faster fall above 10 TeV. 
The Whipple data fit very well with the above three
scenarios. However, above 5 TeV, both the EBL
corrected fit and the exponential fit differ from the Whipple
fit. Again the EBL fit and the exponential fit differ above 10 TeV and
the former one falls faster than the latter as clearly shown in  Fig. \ref{mrk501tevsed}.
Even though all these fit very well with the Whipple data, 
deviation is obvious in the VHE limit. Again, 
between March 16th and October 1st, 1997, Mrk 501 was in a flaring
state which was monitored in TeV $\gamma$-rays with the HEGRA stereoscopic
system of imaging atmospheric Cherenkov telescopes (IACTs). During this
long outburst period (a total exposure time of 110 h) more than 38,000 TeV photons were
detected and the energy spectrum of the source was above 10
TeV. A time-averaged
energy spectrum of this observation period was fitted with a power-law
accompanied with an exponential cutoff \cite{Aharonian:1999vy}. The same data was
reanalyzed with an improved energy resolution\cite{Aharonian:2000xr} and found that except
for the highest energy, the two analysis were in very good agreement. 
At the highest energy the spectrum was found to be much steeper
than the conventional analysis. In Fig. \ref{mrk501tevsed}, along with the 2009 flare
spectrum, we have also shown the conventional analysis and the
improved energy resolution analysis of the 1997 outburst observed by
HEGRA telescopes. The photohadronic model of ref.\cite{Sahu:2016mww} 
fits very well with the reanalysis result of 1997 flare data beyond 10 TeV.
The entire SED (from low to VHE) is shown in Fig. \ref{mrk501sed} and
the intrinsic  flux (red curve in Fig. \ref{mrk501tevsed}) is plotted
to show the EBL contribution.

By comparing different time scales and the Eddington luminosity (as
done for Mrk 421), the optical depth is in the range 
 $0.04 < \tau_{p\gamma} <
0.13$ and this corresponds to the range of photon density in the
inner jet region as $1.5\times 10^{10}\,
cm^{-3} < n'_{\gamma,f} < 5.1\times 10^{10} cm^{-3}$. 
Due to the adiabatic expansion of the inner jet,
the photon density will be reduced to $n'_{\gamma}$ and also the
optical depth $\tau_{p\gamma} \ll\, 1$. This will drastically reduce the $\Delta$-resonance
production efficiency from the $p\gamma$ process. 

\subsection{1ES1959+650}

The AGN 1ES 1959+650 ($z=0.047$)  was first detected in the Einstein IPC Slew Survey\cite{Elvis:1992}
and classified as a HBL subclass, based on its
X-ray to radio flux ratio\cite{Schachter:1993} with a luminosity distance of
$d_L=210$ Mpc and the mass of the central black hole is
estimated to be $\sim 1.5\times 10^8 M_{\odot}$. 
In 1998, VHE gamma-ray from 1ES 1959+650 was observed by the
Seven Telescope Array in Utah
and later on other observations were also reported. In May 2002, 1ES 1959+650  had a
strong TeV outburst which was observed by Whipple\cite{Holder:2002ru} and HEGRA
experiments\cite{Aharonian:2003be} as well as in the X-ray range by RXTE experiments. The
X-ray flux smoothly declined  throughout the following month. However,
during this smooth decline period, a second TeV flare was observed after few
days (on 4th of June) of the initial one without a X-ray
counterpart\cite{Krawczynski:2003fq}. As this flare was not
accompanied by low energy counterparts, it is called {\it orphan} flare.
So the observation of
the orphan flare in 1ES 1959+650  is in striking disagreement with the predictions
of the leptonic models thus challenging the SSC interpretation of the
TeV emission. Similar behavior also observed in the flare of April 2004 from
Markarian 421.
Non observation of a significant X-ray activity could naturally be
interpreted by the suppression of electron acceleration and inverse
Compton scattering as production mechanism for very high energy (VHE) gamma rays in favor
of alternative scenarios. To explain the orphan flare,
A hadronic synchrotron mirror model was
proposed by B\"ottcher\cite{Bottcher:2004qs} to explain this orphan
TeV flare  from 1ES1959+650\cite{Reimer:2005sj}. In
this model, the flare is explained through the decay of neutral pions
to gamma rays when the former are produced due to the interaction of
high energy cosmic ray (HECR) protons with the primary synchrotron photons that have
been reflected off clouds located at a few pc above the accretion
disk. These photons are blue shifted in the jet frame so that there
will be substantial decrease in the HECR proton energy to overcome the
threshold for $\Delta$-resonance and,  at the same time, it is an
alternative to the standard scenario where HECR protons interact with
the synchrotron photons, where one needs HECR protons to be Fermi
accelerated to very high energy. However, how efficiently these photons will be reflected from the
cloud is rather unclear. 

The SED of the 1ES1959+650 is fitted quite well with the leptonic one-zone
synchrotron and SSC model\cite{Tagliaferri:2008qk,Gutierrez:2006ak}. 
In all these
models, although the blob size differ by about 1 to 2 orders of
magnitudes ($1.4\times 10^{14}\,{\rm cm} \le R'_b \le 1.4\times
10^{16}\,{\rm cm})$,
the bulk Lorentz factor $\Gamma$ and ${\cal D}$ are almost the same
($18 \le \Gamma\simeq {\cal D}\le 20$). 
The multiwavelength observation of  1ES
1959+650 was performed in May, 2006 and the SED fitted with the above one-zone
model by Tagliaferri et. al\cite{Tagliaferri:2008qk},
for which the parameters used are
$\Gamma\simeq {\cal D}=18$, $R'_b =7.3\times 10^{15}$ cm and $B'=0.25$
G. To explain the VHE spectrum of the orphan flare in photohadronic
model\cite{Sahu:2013ixa}, $\Gamma=18$ is also used.

%%%%%%%%%%%%%%%%%%%%%%%
\begin{figure}[t!]%fig1
\vspace{0.3cm}
{\centering
\resizebox*{1.0\textwidth}{0.6\textheight}
%{\includegraphics{blazar_spectrum_figure2v2.eps}}
{\includegraphics{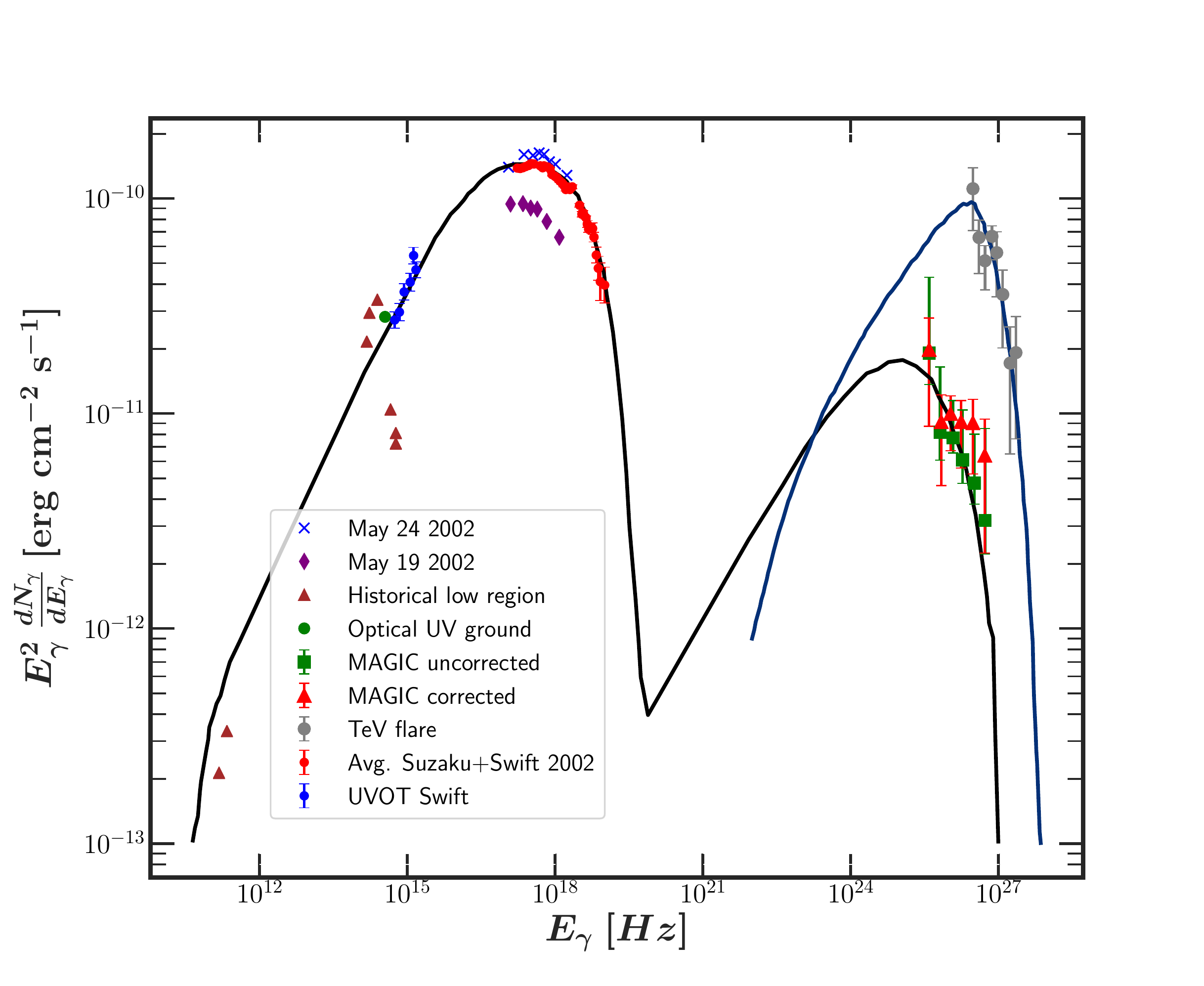}}
\par}
\caption{The multiwavelength SED  measured at the end of 2006 May with
 other historical data from ref.\cite{Tagliaferri:2008qk} are shown.
Different symbols are observations with 
  different sources marked within the box, and the curves are model fits:
  The curve  (from low energy to high energy) is synchrotron + SSC fit from Tagliaferri
   et. al.\cite{Tagliaferri:2008qk}, while 
    the second curve (extreme right) is the photohadronic fit to the
    flare data.
%The X-ray spectra are the single-power-law fit taken
% from Ref. \cite{Krawczynski:2003fq}.
%(See the electronic edition of the Journal for a color version of this figure).
}
\label{1es1959sed}
\end{figure}
%%%%%%%%%%%%%%%%%%%%%%%%%%%
%%%%%%%%%%%%%%%%%%%%%%%
\begin{figure}[t!]%fig1
%\vspace{0.3cm}
{\centering
\resizebox*{1.0\textwidth}{0.4\textheight}
%{\includegraphics{blazar_spectrum_figure2v2.eps}}
{\includegraphics{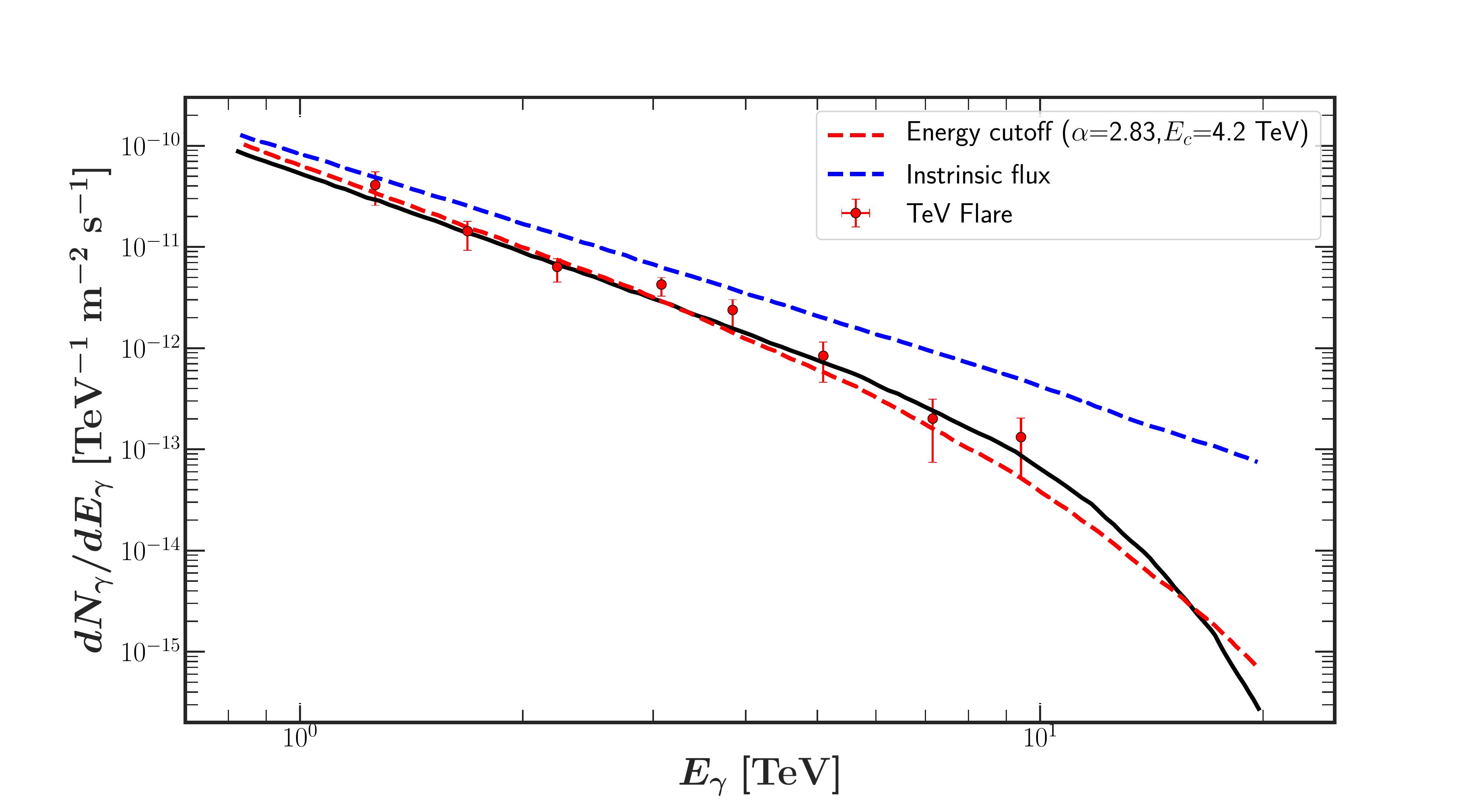}}
\par}
\caption{The flare data are fitted with power-law with exponential
  decay (red dotted curve) and power-law with EBL correction (black
  curve). The blue dotted curve is the intrinsic flux.
%The X-ray spectra are the single-power-law fit taken
% from Ref. \cite{Krawczynski:2003fq}.
%(See the electronic edition of the Journal for a color version of this figure).
}
\label{1es1959tevsed}
\end{figure}
%%%%%%%%%%%%%%%%%%%%%%%%%%%

The observed flare energy was in the 
range $1.26\, {\rm TeV} (3.05\times 10^{26} {\rm Hz})\lesssim E_{\gamma} \lesssim 9.4\, {\rm
  TeV} (2.3\times 10^{27} {\rm Hz})$. Gamma-ray in this energy range is
produced when protons in the energy range $12\, {\rm TeV} \leq E_{p} \leq 94\, {\rm TeV}$
collide with the background photons in the energy interval 
$7.5\, {\rm MeV} (1.8\times 10^{21}\, {\rm Hz})\geq
\epsilon_{\gamma} \geq 1\, {\rm MeV} (2.4\times 10^{20}\, {\rm
  Hz})$. The range of $\epsilon_{\gamma}$ lies exactly in the low energy tail of
the SSC photons as shown in Fig. \ref{1es1959sed},
calculated using the one-zone leptonic model and flux in this region
also has a power-law behavior $\Phi_{SSC}\propto \epsilon^{\beta}_{\gamma}$. As discussed for Mrk
421 and Mrk 501, we can relate the photon density in the inner region
to the outer region and calculate the observed high energy flux.

Best fit to the observed VHE spectrum is obtained with the values of $\alpha=2.83$ and $E_c=4.2$
TeV\cite{Aharonian:2003be} for power-law with
exponential cut-off. The $\gamma$-ray cut-off
energy of 4.2 TeV corresponds to $E_{p,c}=42$ TeV and above the cut-off
energy the flux decreases. 
For power-law with EBL correction, the best fit is obtained for
$\alpha=2.8$ and $A_{\gamma}=87$ (black curve) using 
Eq. (\ref{modifiedsed}). For comparison, both the
exponential and the EBL fits are shown in Fig. \ref{1es1959tevsed}. The
exponential fit falls faster than the EBL fit above 16 TeV.
The time averaged TeV energy spectrum (above 1.4 TeV) of the flaring state of the 1ES
1959+650 was well fitted with pure power-law by the HEGRA
collaboration and the power-law
spectral index is $\alpha=2.83\pm 0.14_{\rm stat} \pm 0.08_{\rm
  sys}$ or by a power-law with an exponential cut-off at
$E_c=(4.2^{+0.8}_{-0.6{\rm stat}}\pm 0.9_{\rm sys})$ TeV and a
spectral index of $1.83\pm0.15_{\rm stat}\pm 0.08_{\rm
  sys}$\cite{Aharonian:2003be,Daniel:2005rv}. 

In Fig. \ref{1es1959sed}, the flare data is shown along with the
fit. Normally, it is observed that the flux increases
for  $E_{\gamma} < 1.2$ TeV due to the  high proton flux in this
energy range. However, in order not to violate the Eddington
luminosity, the proton energy spectrum must break to a harder index
(e. g. $\alpha \sim 2.3$) below 12 TeV. So a break in proton spectrum is introduced 
at $E_{p,b}\sim 12$ TeV below which $\alpha=2.3$ and above this energy
$\alpha=2.8$ so that the gamma-ray flux falls below 12 TeV.

\section{General Remark}

A general discussion of shortcomings in different models is discussed here.
In proton synchrotron models\cite{Cao:2014nia},  the interaction of high energy protons with the
synchrotron photons in the jet can produce $\gamma$-rays from $\pi^0$
decay and can explain the multi-TeV emission from blazars.
Also Zdziarski et al.\cite{Zdziarski:2015rsa,Pjanka:2016ylv} have used the hadronic model to explain the
broad band spectra of radio-loud AGN. These scenarios
require super Eddington luminosity in protons ($\sim 10^6$ times the
Eddington flux ) to explain the
multi-TeV emission. Also, synchrotron emission from the 
ultra high energy protons in the jet magnetic field can 
explain the VHE $\gamma$-ray SED\cite{Mucke:2000rn} but 
strong magnetic field at the emission site is a necessary requirement
which is probably not there.
In an alternative scenario, 
ultra high energy protons escaping from the jet
region produce VHE photons by interacting with the cosmic microwave
background (CMB) photons and/or EBL which avoids the absorption in
the inner jet region\cite{Essey:2009zg}. This
explains the transparency 
of the universe to VHE $\gamma$-rays due to their proximity 
to the Earth compared to the one produced in the source which travels
a longer distance. Also the TeV
spectrum is independent of the intrinsic spectrum but depends on the
output of the high energy cosmic rays in the source. This model fits
very well to multi-TeV spectra from many sources
\cite{Essey:2010er,Essey:2009ju, Prosekin:2012ne,Essey:2013kma}. However,
in this scenario, it is assumed that the source produces VHE protons with energies
$10^{17}-10^{19}$ eV (this is true with most of the hadronic models) and a weak extragalactic magnetic field in the
range $10^{-17}\, G < B < 10^{-14}\, G$ is needed. So far we have not
observed either high energy cosmic ray event or high energy neutrino
event from nearby HBLs (e.g. Mrk 401, Mrk 501
etc.) when they were flaring and also the extragalactic magnetic field
seems to be weak. 
%In an alternative scenario, 
%ultra high energy protons and neutrons \textbf{escaping} from the jet
%region produce VHE photons by \textbf{interacting} with the cosmic microwave
%background (CMB) photons and/or EBL which avoids the absorption in
%the inner jet region\cite{Essey:2009zg}. 
However, in the photohadronic model discussed above,
$p\gamma\rightarrow \Delta$ process takes place in the hidden inner jet
region where the  photon density is order of magnitude higher than the
normal jet and overcomes the super-Eddington energy budget. Also the
energy of the Fermi accelerated proton is $E_p=10 \, E_{\gamma}$ which
can be achieved easily in the jet. 

Normally it is difficult to explain the GeV-TeV emission from HBLs
through one-zone leptonic model. On the other hand, multi-zone leptonic models overcome
this problem and explain the multi-TeV data but one has to increase
the number of parameters.
A power-law with an exponential cut-off energy $E_c$ is the conventional method
used in the literature to explain the exponential fall of the
multi-TeV flare from flaring blazars where the free parameter $E_c$ 
depends on some  unknown mechanism.  The EBL correction does the same
job as the exponential decay factor. So the inclusion of EBL effect
eliminates the necessity of the extra parameter $E_c$.

The high energy protons will be accompanied by high energy electrons and these
electrons will emit synchrotron photons in the magnetic
field of the jet. The energy range of the photons emitted lie in between the high
energy end of the synchrotron spectrum and the low energy tail of 
the SSC spectrum, thus may not be observed due to their low flux in
this region. The secondary charged particles $e^{\pm}$,
$\pi^{\pm}$, $\mu^{\pm}$ produced within the jet will also emit
synchrotron photons and their energy will be much smaller than the
synchrotron photons produced from the primary electrons accompanying
the high energy protons. So even though the charged particles emit
synchrotron radiation, their contribution will be small.

The problem with the photohadronic model is that, the gamma-rays are
produced from the pion decay within the inner jet region where photon
density is very high compered to outer region. So
$\gamma\gamma\rightarrow e^{+}e^-$ should be efficient, which will
attenuate the propagation of multi-TeV gamma-rays from the production
site. However, observation of multi-TeV gamma-ray events from many
flaring HBLs observed by MAGIC, VERITAS, HESS telescopes guarantees
that gamma-rays do escape from the jet without undergoing pair
production. So the pair production process can be constrained.
The photohadronic model works well for high energy gamma-rays (above $\sim
100$ GeV). So in the lower energy regime, it is the leptonic model which
contributes to the multiwavelength SED. In principle both should
coincide at some intermediate energy.

\section{Summary}

One-zone leptonic model is very successful to explain the synchrotron
and SSC peaks of AGN. But this model has difficulties to explain the
multi-TeV emission/flaring from many HBLs. In this article we review
the photohadronic model which is very successful in explaining the
multi-TeV flare data from the nearest HBLs Mrk 421, Mrk 501 and
1ES1959+650. As high energy gamma-ray is attenuated by the EBL, it is
necessary to account for it which is considered here and compared with
the exponential cut-off scenario and other power-law fits. The main
objective here is to understand the flaring from nearby HBLs so that
the mechanism to produced intrinsic flux can be understood well and
also it can be useful to understand the flaring from far off HBLs. 
%In this context we have discussed here some flaring from the HBLs Mrk
%421, Mrk 501 and 1ES1959+650 and found that photohadronic model
%explains very well. 
This model uses the leptonic model parameters which explains the first
two peaks of the AGN very well as input. In the photohadronic scenario,
the observed flux is proportional to the SSC flux $\Phi_{SSC}$ which
is a power-law, $\Phi_{SSC}\propto E^{-\beta}_{\gamma}$ in the lower
tail region. So measurement of SSC
flux in the low energy tail region is required. But, most of the cases
this region of the SED is not observed/measured due to technical
difficulties. Hence, model calculations are used for it and different
models give slightly different $\beta$. As the
photohadronic model depends on $\Phi_{SSC}$, simultaneous multiwavelength
observation of the object during the flaring is important and in few
cases it has already been done. In future, accurate measurement of EBL contribution and observation of
HBLs above 30 TeV energy range will definitely constrain severely different
models discussed above.

 %\acknowledgments

I thank my collaborators Salvador Miranda, Alberto Rosales de Le\'{o}n,
Shigehiro Nagataki, Vladimir Y\'{a}\~{n}ez and Virendra Gupta for many 
simulating discussions. I am thankful to Carlos Lopez Fortin for his help
in drafting the article. The work of S. S. is
partially supported by DGAPA-UNAM (Mexico) Project No. IN103019.

\end{document}